%% file: main_file.tex
\RequirePackage{tikz}

\documentclass[pdflatex,sn-basic]{sn-jnl}

\usepackage[utf8]{inputenc}
\usepackage{textgreek}
\usepackage{subcaption}

\usetikzlibrary{shapes,shapes.geometric,arrows,fit,calc,positioning,automata,}

\jyear{2021}%

\theoremstyle{thmstyleone}%
\theoremstyle{thmstyletwo}%

\theoremstyle{thmstylethree}%

\newcommand\myeq{\mathrel{\overset{\makebox[0pt]{\mbox{\normalfont\tiny\sffamily def}}}{=}}}

\begin{document}

\title[ ]{Link Count Data-driven Static Traffic Assignment Models Through Network Modularity Partitioning}


\author*[1,2]{\fnm{Alexander} \sur{Roocroft}}\email{aroocroft1@sheffield.ac.uk}

\author[1]{\fnm{Giuliano} \sur{Punzo}}

\author[2]{\fnm{Muhamad Azfar} \sur{Ramli}}



\affil*[1]{\orgdiv{ACSE Department}, \orgname{University of Sheffield}, \orgaddress{\city{Sheffield}, \country{UK}}}

\affil[2]{\orgdiv{Institute of High Performance Computing}, \orgname{A*STAR}, \orgaddress{\country{Singapore}}}




\abstract{Accurate static traffic assignment models are important tools for the assessment of strategic transportation policies. In this article we present a novel approach to partition road networks through network modularity to produce data-driven static traffic assignment models from loop detector data on large road systems. The use of partitioning allows the estimation of the key model input of Origin-Destination demand matrices from flow counts alone. Previous network tomography-based demand estimation techniques have been limited by the network size. The amount of partitioning changes the Origin-Destination estimation optimisation problems to different levels of computational difficulty. Different approaches to utilising the partitioning were tested, one which degenerated the road network to the scale of the partitions and others which left the network intact. Applied to a subnetwork of England's Strategic Road Network and other test networks, our results for the degenerate case showed flow and travel time errors are reasonable with a small amount of degeneration. The results for the non-degenerate cases showed that similar errors in model prediction with lower computation requirements can be obtained when using large partitions compared with the non-partitioned case. This work could be used to improve the effectiveness of national road systems planning and infrastructure models.}




\maketitle

\input{Intro1}
\input{MethodBackground1}
\input{MethodNetworkSimple}

\input{Results}
\input{Conclusion1}
\input{Appendix}

\bibliography{bibliography}



\end{document}

%% file: Intro1.tex
\section{Introduction}
Public investment to alleviate congestion on national road networks attracts much scrutiny due to the high costs involved and the essential nature of key infrastructure. Having accurate models of road traffic to allow policy makers to undertake long-term planning are therefore necessary. Static Traffic Assignment (TA) models are frequently used for strategic transportation planning within travel demand models (\cite{DfT2022}). Much of current research focuses on dynamic TA which can model congestion more accurately, however static TA still has a specific use for the economic appraisal of long-term future changes to traffic patterns at the entire network level (\cite{Patil2021,Tsanakas2020}).

Origin-Destination (O-D) demand estimation is a key challenge for static TA models and road transportation planning. O-D demand matrices represent the number of trips taken by drivers between distinct origins and destinations on the road network within a specific analysis time period (\cite{Abrahamsson1998}). In the literature there are a range of different approaches for their estimation (\cite{Bera2011}).

An established way of obtaining O-D matrices is through manual surveys of road users. However, these can be expensive and laborious, having low sample rates leading to high sampling bias risk and missed movements (\cite{Hazelton2000}). As an alternative in recent years there has been interest in new forms of historic trip data which provide information on driver trajectories from sources such as mobile phone GSM, GPS and Automatic Number Plate Recognition (ANPR) (\cite{Cvetek2021,Landmark2021,Liao2022}). However, these types of data have issues relating to privacy and integration into the road network which limits their accessibility for data-driven modelling (\cite{Mahajan2021}). Further approaches include utilising zone-based activity and socio-economic data to simulate approximate theoretical demands (\cite{Horni2016,Ren2014}), and toll gate data in closed highway systems (\cite{Zeng2021}).

In many countries, inductive loops under the main strategic roads are used to monitor traffic. Often this data is publicly available and does not entail privacy concerns (\cite{GraphHopper}). However, inductive loops do not provide any information on the routes drivers take. Techniques in the literature exist which can use flow count data from loop detectors to estimate O-D demand without the additional need for survey or historic trip data.

Attempting to estimate the O-D matrix solely from mean traffic flows entails problems relating to identifiability as the number of link flow counts is less than the number of O-D demand pairs to be estimated so it is difficult to know which vehicles on a road are travelling between which O-D nodes (\cite{Hazelton2003}). 

Network tomography-based approaches such as (\cite{Hazelton2000,Vardi1996,Lo1996,Dey2020}) attempt to use the stochastic nature of traffic counts to estimate O-D demands using multiple samples of link flows on the network for the estimation time period. Assuming the Poisson distribution of demands and a non-congested network, the Generalised Least Squares (GLS) as formulated in \cite{Hazelton2003} is a practical version of this approach which has been applied to real world highway networks in static TA models (\cite{Zhang2018}). Although its assumptions may be strong (\cite{Tebaldi1998}), due to its relatively lower computational requirements compared to the other network tomography-based approaches, the GLS is useful for gaining a prior  matrix to be subsequently refined to include the effects of congestion through O-D adjustment algorithms (\cite{Spiess1990,Lundgren2008}). Other related flow count techniques are reported to have superior accuracy, however they require additional data sources such as privacy sensitive ANPR (\cite{RostamiNasab2020,Yang2017,ParryHazelton2012}).

GLS and network tomography-based techniques in general have difficulties working with large network sizes due to high space and time complexity in the involved processes (\cite{Brander1996}). Previous real-world applications of GLS have been limited to 34 node road networks with routes between O-D pairs limited to one (\cite{Zhang2018}). Other network tomography-based approaches have been applied to smaller sized road networks (\cite{Dey2020,Hazelton2001}).

In this article, we propose a novel method to apply link flow count O-D estimation to large-scale real-world road networks. This is done by partitioning the network into communities of smaller subnetworks to apply  estimation to. We carry out an analysis of how partitioning a road network into a range of sizes affects accuracy and computational requirements.

Our partitioning approach uses community detection. Many networks representing complex systems contain a modular structure where the nodes cluster into communities of relative high density of connections with fewer connections between (\cite{Traag2019}).
A well-known performance measure to detect such community structure is network modularity (\cite{Fortunato2010}). One of the most used algorithms to evaluate modularity, which is an NP-complete problem (\cite{Brandes2006,Leeuwen2019}), is the Louvain algorithm which allows the evaluation of a hierarchy of community partitions to be made (\cite{Blondel2008}). A resolution parameter can determine the size of clusters that are identified. Applied to a road network, this can group areas of the network into clusters which are internally well-connected and externally less strongly. Basing community detection and the resulting partitioning on modularity utilises the network distance and not geographic distance between  pairs which can be different. The grouping of nodes closer together on the network benefits the GLS estimation as the method does not account for geography constraints explicitly (\cite{Dey2020}). 

Previously, modularity and Louvain have been used to investigate high-level spatial and temporal patterns in travel demand when the demand is known, finding a strong relation between demand and geographic closeness of O-D pairs (\cite{Leeuwen2019}). This provides evidence that the structure of travel demand could work with partitioned  estimation.

Other works in transport literature have partitioned road networks with different approaches, utilising it for microscopic simulation (\cite{Ahmed2016}), macroscopic fundamental diagrams (\cite{Hirabayashi2019,Lin2020}), and traffic management through travel speed correlation (\cite{Yu2021}). As far as we know, previous research has not used partitioning the road network via network modularity for link-count demand estimation within static TA.

Our work develops several ways of applying partitioning to the  estimation problem. The partitions can be the basis of reducing the road network down to a smaller, degenerated network with single nodes representing each community. Such a model could be integrated into infrastructure models such as NISMOD in the UK (\cite{Blainey2019}) which work at the scale of large urban areas but lack accurate treatment of traffic modelling. The partitions are also used within non-degenerate approaches which preserve the road network in full but utilise the different scales of analysis, internal and external to the partitions, to estimate a full network demand matrix with increased agility.

Standard validation techniques are often inadequate to assess the effects of the partitioning on the  estimates (\cite{Dey2020}). Comparing the estimated  matrix to another validation data source, such as historic trips, is problematic as that is still only a sample of the  movements. It is impractical to account for all the movements on a large-scale road network for a ground-truth matrix. For this reason the validation of the results is done via the relative accuracy, predicting the flow and travel times through the user-equilibrium flow pattern of a derived static TA model.

To test this approach primarily we use link count data from the England Strategic Road Network (SRN), a large real-world non-closed highway network suitable as a case study. This new technique is applied to a sample subnetwork connecting major metropolitan areas in England, using traffic flow count data obtained from the Motorway Incident Detection and Automatic Signalling (MIDAS) system used by National Highways (England) on the National Traffic Information Service (NTIS) model.

\subsection{Summary of contribution}
In this work, we propose a novel integrated and scalable method to obtain O-D estimations for large real-world highway networks and evaluate its performance producing accurate user-equilibrium flow patterns with static TA models. We do this by using network modularity as a basis for dividing up the road network into partitioned subnetworks to reduce the computational difficulty of the O-D estimation problem. This new technique is applied to a large portion of England's SRN. It is demonstrated that the incorporation of partitioned O-D estimation within user-equilibrium flow pattern calculation has the effect of enabling reasonable estimates of the predicted flows and travel times compared to the unpartitioned case while greatly reducing the computational requirements. It is shown in the results that non-degenerate internal-only and internal-external combined approaches with large partitions leads to the best accuracy.

The primary contributions of our work are summarized as:
\begin{itemize}
\item  A new method of producing O-D matrices from flow counts is proposed which utilises network modularity to determine the optimal way to partition the network effectively and automatically.
\item  The new method is applied in the calculation of user-equilibrium flow patterns solely from loop detector data on large scale real-world networks without the current size limitations of similar existing O-D estimation techniques.
\item Different approaches to utilising the partitioning are investigated: one degenerates the network based on the partitioning; others use the partitioning to focus on estimating the prior matrix from the internal and/or external movements of the partitioned nodes. It is found that using within-the-partition internal estimates for the O-D appraisal provide the best accuracy. Including the external between-the-partition estimates can help computation time.
\end{itemize}
The overall structure of this paper is summarized as follows. Section 2 describes the overall methodology for creating a full single-class, static TA model using network and loop detector data. In Section 3, the method of network simplification is presented. Section 4 provides a summary description of the MIDAS and NTIS datasets used for the case study. In Section 5 the main results are presented. Lastly, the paper is concluded with a discussion in Section 6, followed by a conclusion in Section 7.

%% file: MethodBackground1.tex
\section{Traffic Assignment Model Description}

\subsection{Preliminaries and notation}\label{sec2}

\subsubsection{Notation}
In this work all the vectors are column vectors. For example, the column vector \textbf{x} is written as $\textbf{x} =\{x_i,...,x_{dim(\textbf{x})}\}$, where dim(\textbf{x}) is the dimension of \textbf{x}. We use "prime" (e.g. \textbf{x}') to denote the transpose of a matrix or vector.  $\mathbb{R}_+$ denotes the set of all non-negative real numbers. Matrix \textbf{Q} $\geq$ \textbf{0} or vector \textbf{x} $\geq$ \textbf{0} indicates that all entries of a matrix \textbf{Q} or vector \textbf{x} are non-negative. Also, $\vert\mathcal{X}\vert$ represents the cardinality of a set $\mathcal{X}$, and $[\![\mathcal{X}]\!]$ is used for the set $\{1,...,\vert\mathcal{X}\vert\}$.

\subsubsection{Network definition}

When applied to the England SRN, the NTIS model edges and nodes are grouped into superedges and supernodes which are used when referring to the simplified topographic representation. Each supernode is a group of NTIS model nodes which comprise motorway junctions. Each superedge is a collection of the NTIS model edges which comprise each carriageway between the junctions. The road network is modelled as a directed graph with a set of supernodes $\mathcal{V}$  and a set of superedges $\mathcal{A}$. The supernodes represent the interchanges of the road system and the superedges are the roads connecting between them.   The model assumes the graph is strongly connected and is defined by the node-edge incidence matrix with $ N \in \{0,1,-1\}^{(\vert\mathcal{V}\vert\times\vert\mathcal{A}\vert)}$. On road networks in general and the England SRN in particular there is a path between all pairs of supernodes so the assumption is valid.

The demand for movement between O-D pairs on the network is represented by $d^\textbf{w}\geq0$ with \textbf{w} $=(w_s,w_t)$ the O-D pair of supernodes such that $\mathcal{W} =\{\textbf{w}_i :\textbf{w}_i =(w_{si},w_{ti}),i=1,...,\vert\mathcal{W}\vert\}$. $\mathbf{d}^\mathbf{w}\in\mathbb{R}^{\vert\mathcal{V}\vert}$ is a vector with all zeros except for a $-d$\textsuperscript{\textbf{w}} for supernodes $w_s$ and a $d$\textsuperscript{\textbf{w}}  for supernodes $w_t$.
For demand estimation, the O-D demand matrix is denoted in vectorised form as $\textbf{g}=(g_i; i\in[\![\mathcal{W}]\!])$ with each $g_i$ equivalent to a $d^{\textbf{w}}$. $\mathcal{R}_i$  is the index set of simple routes (without cycles) connecting O-D pair $i\in[\![\mathcal{W}]\!]$.

Let \textbf{x} be the vector of the total edge flow $x_a$ on superedge $a\in\mathcal{A}$. Then the set of feasible flow vectors \( \mathcal{F} \) is defined by:
\begin{equation*}
 \mathcal{F} \myeq \{\textbf{x}: \exists\textbf{x\textsuperscript{w}}\in\mathbb{R}_+^{\vert\mathcal{A}\vert} ~s.t. \, \textbf{x} = \sum_{\textbf{w}\in \mathcal{W}} \textbf{x}^\textbf{w},\mathbf{Nx}\textsuperscript{\textbf{w}}=\textbf{d}\textsuperscript{w}, \forall{}\textbf{w}\in{\mathcal{W}} \}
\end{equation*}

where $x^\textbf{w}$ indicates the flow vector attributed to O-D pair \textbf{w}. This implies that the total flow vector \textbf{x} is consistent with the demands \textbf{d}\textsuperscript{w} between all O-D pairs.

The methods described in the following sections use different days of flow data on the network. They are seen as “snapshots” of the network at different points in time, with $\vert\mathcal{J}\vert$ samples of the superedge flow vector \textbf{x}. $j$ $\in[\![\mathcal{J}]\!]$ where $j$ is the index of different snapshots of the network with corresponding average time-bin hourly flows. 

A collection of the network variables is provided in Table \ref{tab:Definitions}.

\begin{table}[t]
	\caption{Notation for Network Definition}
	\begin{center}
		\begin{tabular}{l l l}
            \hline
             Symbol & Definition\\\hline
			$\mathcal{V}$ & Set of Supernodes \\
			$\mathcal{A}$ & Set of Superedges \\
            $\mathcal{W}$ & Set of O-D Pairs \\
            $\mathcal{F}$ & Set of Feasible Flow Vectors\\
            $\mathcal{R}_i$ & Set of Simple Routes for O-D pair $i$\\
            $\mathcal{J}$ & Set of Time-bin Average Flow Vector Samples \\
            $N$      & Node-edge Incidence Matrix\\
            \textbf{g} & O-D Demand Vector \\
            $x_a$ & Flow on Edge $a\in\mathcal{A}$\\\hline
            
		\end{tabular}
	\end{center}
\label{tab:Definitions}
\end{table}

\subsection{Congestion functions}
Accurate congestion functions are key to TA models as they connect the travel time $t_a$ to the vehicle flows $x_a$ on edge $a\in\mathcal{A}$. In the network model they take the form:\\
\begin{equation} \label{eq:3}
\centering
 	t_a(x_a)=t_a^0\,g\left(\frac{x_a}{m_a}\right)~,
\end{equation}\\
where $t_a^0$ is the free-flow travel time of an edge $a \in \mathcal{A}$ and $g(\cdot)$ is a strictly increasing and continuously differentiable function dependent on the flow $x_a$ divided by the flow capacity $m_a$  of that edge $a \in \mathcal{A}$. 

 The BPR equation is consistent with Eq. \ref{eq:3} and is widely used in TA models (\cite{DeGrange2017,Youn2008}). In its more general form it is:
\begin{equation} \label{eq:BPR}
	t_a=t_a^0\left(1+\alpha{(\frac{x_a}{{{m}_a}})}^\beta\right)~,
\end{equation}
where the values of \textalpha \, and \textbeta\, are coefficients commonly taken as 0.15 and 4, respectively (\cite{Sheffi1985}). In this work we use this form of BPR and coefficients for all superedges.

As in \cite{Dervisoglu2009}, we use the maximum of the observed per-minute flows on a superedge as its capacity. The NTIS provided values of capacity are used for edges without sufficient congestion data for this estimation. The free-flow travel time $t^{0}_a$ is obtained by taking the 95\textsuperscript{th} percentile of the observed per-minute speeds (\cite{Casey2020,Silvano2020}) as the free-flow speed then converting it to the travel time through the superedge length.

For free-flow travel time and capacity estimation, outliers in the recorded traffic data are reduced through a 10-min rolling mean applied to the per-minute observations.

\subsection{Estimating the O-D demand matrix}
We use the GLS method together with the Bi-Level optimisation problem (BiLev) algorithm to estimate the O-D demand matrix. 

The GLS method assumes the edges are uncongested so that for each O-D pair the route choice probabilities are independent of the traffic flows. It obtains the estimated vectorised O-D demand matrix \textbf{g} ($\geq$\textbf{0}) through the following optimisation problem (see \cite{Hazelton2000} for details):
\begin{align} \nonumber \label{eq:GLS}
&\min_{\textbf{P}\geq\textbf{0},\textbf{g}\geq\textbf{0}}\quad\sum_{j=1}^{\vert\mathcal{J}\vert}(\textbf{x}^j-\textbf{BP'g})'S^{-1}(\textbf{x}^j-\textbf{BP'g})\\
&s.t.\quad p_{ir}=0\;\forall\;(i,r) \in \{(i,r):r\not\in\mathcal{R}_i\}\\\nonumber
&\quad\quad\textbf{P1}=\textbf{1}
\end{align} 
Where $\boldsymbol{P}=[p_{ir}]$ is the route choice probability matrix, $\boldsymbol{1}$ is a vector of ones and $S$ is the sample covariance matrix for flows.

For all $a \in \mathcal{A}$ , $r \in \mathcal{R}_i$, and $i\in{1,...,\vert\mathcal{W}\vert}$, the edge-route incidence matrix entry $B_{ra}^i$ is 1 if route $r \in\mathcal{R}_i$ uses edge \textit{a}, or 0 otherwise. 
We find the feasible routes for each O-D node pair using Yen's multiple shortest paths algorithm (\cite{Brander1996}) and use them to create the edge-route incidence matrix \textbf{B}. We limit feasible routes to the four shortest routes by distance as it is common for the majority of the flows to use 3 or 4 choices (\cite{Bonsall1997}).

\subsection{O-D matrix congestion adjustment}

To account for the effects of congestion and improve the accuracy of the initial prior demand estimate \textbf{g\textsuperscript{0}}, the congestion functions can be used to find an improved solution through a gradient-based BiLev algorithm. With the observed flow vector denoted by \textbf{\~{x}} = ($\Tilde{x}_a; a \in \mathcal{A}$) and the estimated user-optimal flow vector \textbf{x}(\textbf{g}) for any feasible non-negative vector \textbf{g} ($\geq$\textbf{0}). The problem is expressed through the following objective function:
\begin{align} 
\min_{\textbf{g}\geq\textbf{0}} \quad F(\textbf{g}) =\: \sum_{i\in[\![\mathcal{W}]\!]}(g_i-g_i^0)^2+ \sum_{a\in\mathcal{A}}(x_a(\textbf{g})-\Tilde{x}_a)^2 
\end{align} 
Further details can be found in \cite{Spiess1990} and \cite{Lundgren2008}.

\subsection{Flow pattern calculation}
The predicted user-equilibrium flow pattern can be calculated using the adjusted O-D demand matrix and congestion functions through the Frank-Wolfe algorithm with the following optimisation of the Traffic Assignment Problem (TAP):
\begin{equation}
\min_{\textbf{x} \in \mathcal{F}} \sum_{a \in \mathcal{A}} \int_{0}^{x_a} t_a(s) ds
\label{eq:UserObj}
\end{equation}

The Frank-Wolfe algorithm uses a convergence criterion based on the size of relative gap between consecutive iterations (see \cite{Patriksson1994} for details). In this work a non-dimensional relative gap of $10^{-5}$ is used for the convergence of the edge flows (\cite{Patil2021}). The user-equilibrium flow pattern results from drivers pursuing their selfish best route and throughout this work it is assumed to match the observed flows as commonly done in other works (\cite{Zhang2018,DeGrange2017}). 

%% file: MethodNetworkSimple.tex
\section{Network Simplification}

\subsection{Network partitioning}
Clustering is performed on the topographic representation based on a community partitioning using network modularity via the Louvain algorithm.

Network modularity measures the relative density of edges inside communities compared to the edges outside communities. It is measured with a scale value ranging from -0.5 to 1 (non-modular to fully modular clustering). By achieving the optimal value for modularity (closest to 1) the results should be the best possible grouping of the network nodes.
The Louvain algorithm works by first finding small communities based on optimising modularity on all the nodes locally. Then those small communities are regrouped as single nodes in a condensed graph and the modularity between them is calculated. A change in modularity process is applied to this new network to see if there are increases in modularity from combining the new community partitions. If no increase in modularity occurs then that partition is optimal, otherwise the process of regrouping the nodes of the condensed graph repeats combining the communities further. See \cite{Blondel2008} and \cite{Traag2019} for more details.

The algorithm uses the following definition for modularity:
\begin{equation}
Q= \frac{1}{2m} \sum \left[  A_{ij} - \frac{k_{i}k_{j}}{2m}  \right]\delta(c_i,c_j)
\end{equation}
where $A_{ij}$ is the weight of the edge between nodes $i$ and $j$ taken as the inverse of the edge length; the sum of the weights of the edges attached to node $i$ is represented by $k_i=\sum_{j} A_{ij}$. The \textdelta-function \textdelta($c_{i}$, $c_{j}$) is 1 if $c_i=c_j$ and 0 otherwise, such that $c_i$ is the community to which node \textit{i} is assigned. Also, $m=\frac{1}{2} \sum_{ij} A_{ij}$ is based on the total weight of network edges.

We partition the topographic representation using the inverse of the superedge distances as the network edge weights as opposed to the true road distance. This is so that nodes closer on the topographic network are treated as having a stronger connection. In the process, we replace pairs of parallel edges which have opposite flow directions with undirected edges due to the Louvain implementation used (\cite{python-louvain}). This does not affect the final result due to carriageways being in identical pairs.

With efficiency for large networks, the Louvain algorithm finds different high modularity partitions for chosen resolutions of community detection. The resolution size is a parameter of the algorithm that affects the size of the communities, making it larger leads to a smaller number of partitions being produced with a greater number of nodes inside each one (\cite{Blondel2008}). We vary the size of the resolution over a range to produce partition sizes from unpartitioned (resolution equals zero) to the largest partitions when there are only two separate communities (a resolution value which depends on network size). Not every resolution produces a unique number of communities. We selected the lowest resolution which found each unique number of communities.

Each time the Louvain algorithm is run with the same inputs it can produce a variation on the exact partitioning produced due to randomized cluster initialization (\cite{Leeuwen2019}). As we are primarily using partitioning to find communities of different sizes, control of the exact nodes in each partition is not a great concern.

\begin{figure}[t]
\centering
\begin{subfigure}[h]{.35\textwidth}
\includegraphics[width=\textwidth]{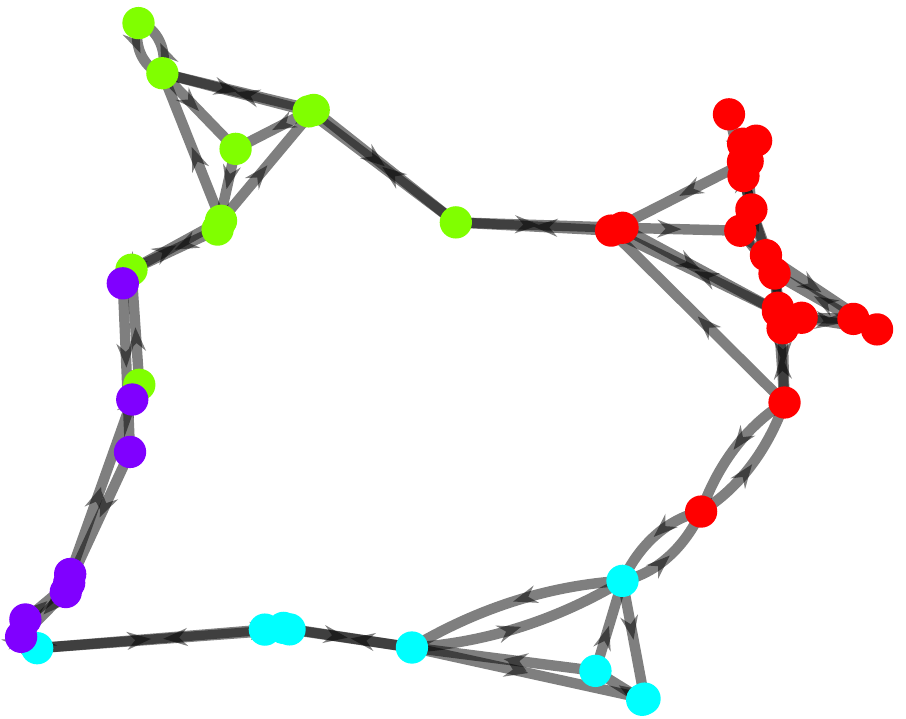}
\end{subfigure}
{\LARGE$\xrightarrow{}$}%
\begin{subfigure}[h]{.35\textwidth}
\includegraphics[width=\textwidth]{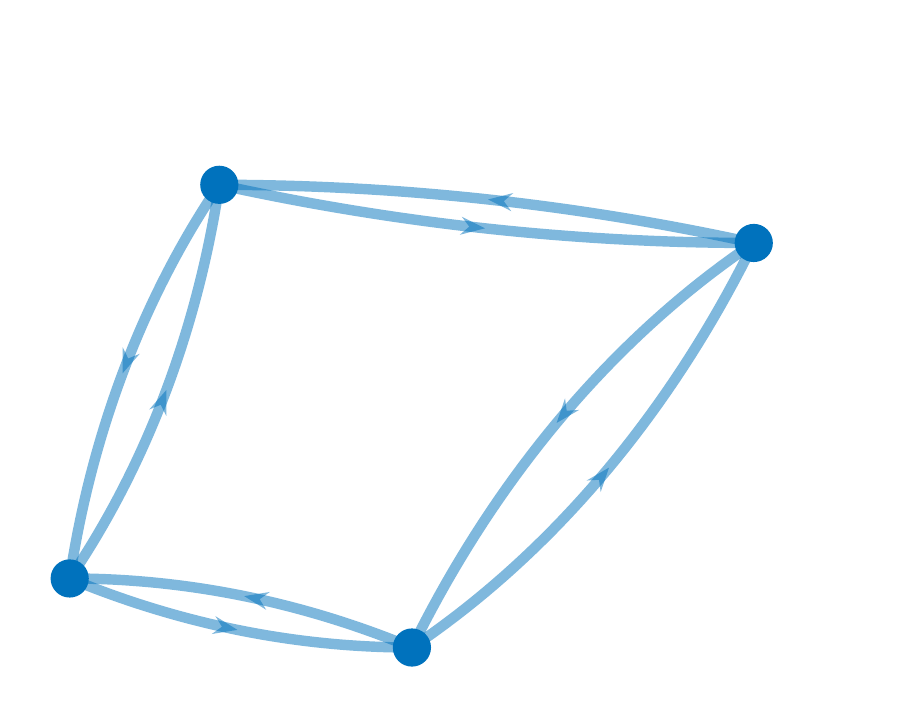}
\end{subfigure}
\caption{Example of community topographic representation after partitioning via Louvain algorithm.}
\label{fig:simplificationExample}
\end{figure}

Once we have produced a result for the given resolution, the new community topographic representation is created from the groupings. The supernodes of each partition are grouped into community supernodes. We use a modified Depth First Search (\cite{Mehlhorn2008}) to find the neighbours of each partition and establish the community superedges of a new community topographic representation. An example of the process can be seen in Figure \ref{fig:simplificationExample}. If multiple superedges connect the partitions then the mean of the superedges weighted by mean flow is used as the community superedge distance and free-flow travel time. The sum of the flows and capacities on the constituent superedges are used as the community superedge flow and capacity, respectively. Because the partitions are adjacent to each other it is often the case that only one superedge forms the community superedge.

\subsection{Partitioned network O-D estimation}
\begin{figure}[t]
\centering
\begin{tikzpicture}[node distance={10mm}, thick, main/.style = {draw, circle}] 
\node[main] (1) {$1$}; 
\node[main] (2) [above right of=1] {$2$};
\node[main] (3) [below right of=1] {$3$}; 
\node[main] (4) [above right of=2, xshift=1.5cm] {$4$};
\node[main] (5) [above right of=4] {$5$}; 
\node[main] (6) [below right of=5] {$6$};
\node[main] (7) [below right of=6, xshift=1.5cm] {$7$};
\node[main] (8) [below right of=7] {$8$}; 
\node[main] (9) [below left of=8] {$9$};
\draw (1) -- (2);
\draw (2) -- (3);
\draw (1) -- (3);
\draw (2) -- (4);
\draw (4) -- (5);
\draw (5) -- (6);
\draw (4) -- (6);
\draw (7) -- (6);
\draw (7) -- (8);
\draw (8) -- (9);
\draw (9) -- (7);
\draw (3) -- (9);

\draw [dashed,red] (1.north) to [out=80,in=190] (2.west);
\draw [dashed,red] (1.south) to [out=280,in=170] (3.west);
\draw [dashed,red] (2.south east) to [out=315,in=45] (3.north east);

\draw [dashed,red] (4.north) to [out=80,in=190] (5.west);
\draw [dashed,red] (5.east) to [out=350,in=100] (6.north);
\draw [dashed,red] (4.south east) to [out=315,in=225] (6.south west);

\draw [dashed,red] (7.east) to [out=350,in=100] (8.north);
\draw [dashed,red] (8.south) to [out=260,in=10] (9.east);
\draw [dashed,red] (7.south west) to [out=225,in=135] (9.north west);

\node (X) [draw=blue, fit= (1) (2) (3), inner sep=0.4cm, 
                dash pattern=, ultra thick, fill opacity=0.2] {};
 \node [yshift=-1.0ex, blue] at (X.south) {A};       
 
\node (Y) [draw=blue, fit= (4) (5) (6), inner sep=0.4cm, 
                dash pattern=, ultra thick, fill opacity=0.2] {};
\node [yshift=-1.0ex, blue] at (Y.south) {B};

\node (Z) [draw=blue, fit= (7) (8) (9), inner sep=0.4cm, 
                dash pattern=, ultra thick, fill opacity=0.2] {};
\node [yshift=-1.0ex, blue] at (Z.south) {C};  

\draw [ultra thick,dash pattern=, blue] (X.north east) -- (Y.west);
\draw [ultra thick,dash pattern=, blue] (Y.east) -- (Z.north west);
\draw [ultra thick,dash pattern=, blue] (Z.west) -- (X.east);

\draw [dashed,teal] (X.80) to [out=70,in=180] (Y.160);
\draw [dashed,teal] (Y.20) to [out=0,in=110] (Z.100);
\draw [dashed,teal] (X.315) to [out=350,in=190] (Z.225);

\end{tikzpicture} 
\caption{Example nine node topographic network (black) partitioned into three communities. Community topographic is in blue. The green dashed lines are the partitions' external O-D movements, the red dashed are the partitions' internal O-D movements.} \label{fig:Example9node}
\end{figure}
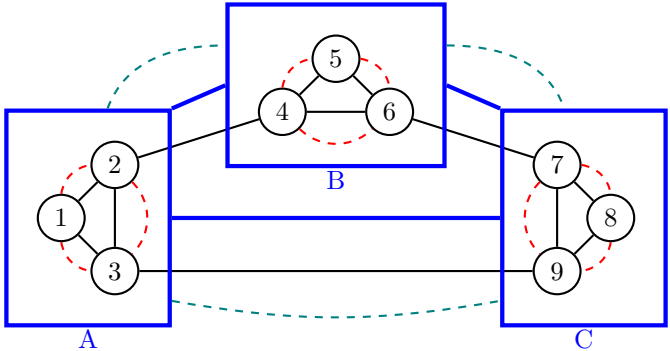

The community topographic representation can be used to obtain estimates of the uncongested O-D demand matrix using the GLS method. We investigate four different ways of utilising the representation for this purpose: (i) degenerate; (ii) non-degenerate internal-only; (ii) non-degenerate external-only; (ii) non-degenerate internal-external combined. Figure \ref{fig:Example9node} shows an illustrative example of a nine node undirected simple graph network to demonstrate the partition grouping with internal and external O-D movements.

\subsubsection*{\textit{(i) Degenerate:}}
In the degenerate O-D estimation, the community topographic representation (Figure \ref{fig:Example9node} - blue graph) is used as a substitute network for original topographic representation (Figure \ref{fig:Example9node} - black graph). The O-D estimation and adjustment are applied to the flows and structure of the community topographic representation and not the original topographic representation. 

In the nine node example, the partitioned community topographic representation is used to produce an O-D estimate, $H_{com}$, for the partitions A, B and C.
\begin{equation*}
\mathbf{H_{com}} = 
\begin{bmatrix}
0 & H_{com}^{AB} & H_{com}^{AC}\\
H_{com}^{BA} & 0 & H_{com}^{BC}\\
H_{com}^{CA} & H_{com}^{CB} & 0 
\end{bmatrix}
\end{equation*}
where each non-zero entry (e.g. $H_{com}^{XY}$) is an estimate of the demand travelling between the pair of partitions (e.g. $X$ and $Y$) based on the link flows of the community superedges (Figure \ref{fig:Example9node} - green dashed lines).

This approach reduces the network size as shown in Figure \ref{fig:simplificationExample}. It loses the detail of individual road junctions but seeks to preserve some of the main network structure. $\mathbf{H_{com}}$ is used within the TA model to produce estimates of flows and travel times between the partitions on the community topographic representation.

\subsubsection*{\textit{(ii) Non-degenerate internal-only:}}
The non-degenerate approaches aim to find an estimate of the demand for each O-D pair of the original topographic representation through breaking down the problem with the simpler community topographic representation.

The internal approach applies O-D estimation to separately estimate demands for the internal O-D pairs of each partition by applying GLS to the flows and structure of that partition's subnetwork (Figure \ref{fig:Example9node} - red dashed lines). For example, for Partition A in the nine node example, we can express a matrix of demands $\mathbf{H_{int}^{A}}$:
\begin{equation*}
\mathbf{H_{int}^{A}} = 
\begin{bmatrix}
0 & H^{12} & H^{13}\\
H^{21} & 0 & H^{23}\\
H^{31} & H^{32} & 0 
\end{bmatrix}
\end{equation*}
where each non-zero entry is an estimate of the demand travelling between the pair of nodes based on the link flows of the topographic representation (Figure \ref{fig:Example9node} - black graph). It follows the same form for other partitions.

For each partition, the O-D values between the internal supernodes will be larger than what would be estimated if the whole unpartitioned network was being used as all the flows are assumed to be going only between the internal supernodes. This is corrected with the help of the O-D adjustment algorithm. 
 
In the non-degenerate internal-only approach, the matrices of demands for each of the partitions are combined into a prior matrix $\mathbf{H}$ by assuming zero demand for the inter-partition O-D pairs. Such that for the nine node example the prior estimate is,
\begin{align*} 
\mathbf{H} & = 
\begin{bmatrix}
\mathbf{H_{int}^{A}} & \mathbf{0} & \mathbf{0}\\
\mathbf{0} & \mathbf{H_{int}^{B}} & \mathbf{0}\\
\mathbf{0} & \mathbf{0} & \mathbf{H_{int}^{C}}
\end{bmatrix}
\end{align*} 
where $\boldsymbol{0}$ is a matrix of zeros the size of the inter-partition O-D pairs.

\subsubsection*{(\textit{iii) Non-degenerate external-only:}}
The non-degenerate external-only approach uses the external partition O-D estimate, $\mathbf{H_{com}}$, obtained from the community topographic representation. The external partition O-D demands are divided equally between the supernodes which comprise the relevant partitions to spread the demand amongst the O-D pairs of the topographic representation (black graph).

 To obtain estimates for the inter-partition demands, the community O-D matrix demands $H_{com}$  are divided by the number of topographic O-D pairs which comprise each partition pair. For example for partition pair AB, the number of nodes in A, $u^A$, is 3 and the number of nodes in B, $u^B$, is 3 so the number of O-D pairs is $u^{AB} = u^A * u^B = 9$. The value for each pair is then $H_{com}^{AB}/9$.
For example, in matrix form, for partition pair AB with $\boldsymbol{1}$ as a column vector of ones the length of the number of nodes in A and B,
\begin{equation*}
\mathbf{\hat{H}_{ext}^{AB}} = 
\frac{H^{AB}_{com}}{u^{AB}}\mathbf{1}\mathbf{1}'
\end{equation*}

External-only assumes zero values for the demands between the O-D pairs internal to the partitions, resulting in the following prior matrix,

\begin{align*} 
\mathbf{H} & = 
\begin{bmatrix}
\mathbf{0} & \mathbf{\hat{H}_{ext}^{AB}} & \mathbf{\hat{H}_{ext}^{AC}} & \\
\mathbf{\hat{H}_{ext}^{BA}} & \mathbf{0} & \mathbf{\hat{H}_{ext}^{BC}} & \\
\mathbf{\hat{H}_{ext}^{CA}} & \mathbf{\hat{H}_{ext}^{CB}} & \mathbf{0} \\
\end{bmatrix}
\end{align*} 
where $\boldsymbol{0}$ is a matrix of zeros the size of the intra-partition O-D pairs.

\subsubsection*{\textit{(iv) Non-degenerate internal-external combined:}} 

In the non-degenerate internal-external combined approach, a prior matrix is formed using both internal and external estimations without any O-D demands assumed zero:

\begin{align*} 
\mathbf{H} & = 
\begin{bmatrix}
\mathbf{H_{int}^{A}} & \mathbf{\hat{H}_{ext}^{AB}} & \mathbf{\hat{H}_{ext}^{AC}} & \\
\mathbf{\hat{H}_{ext}^{BA}} & \mathbf{H_{int}^{B}} & \mathbf{\hat{H}_{ext}^{BC}} & \\
\mathbf{\hat{H}_{ext}^{CA}} & \mathbf{\hat{H}_{ext}^{CB}} & \mathbf{H_{int}^{C}} \\
\end{bmatrix}
\end{align*} 

In all the non-degenerate approaches, we use the prior matrix $\mathbf{H}$ in the O-D adjustment algorithm to produce a final O-D demand matrix which is used in a static TA model for the whole topographic network.

\section{Application on the England Strategic Road Network}

\subsection{Raw dataset description}
The dataset used in this work takes the traffic data obtained through the MIDAS system installed on the main motorways and A-roads of the England SRN. MIDAS mostly records traffic through under-road inductive loops spaced approximately every 500m. The dataset includes the data for the weekdays in September 2018 to June 2019.
The MIDAS system measures speed, flow, occupancy and headway at approximately 7000 sites on the SRN. The data is given on a per-lane basis and aggregated over 1-minute intervals.  It is assumed that the network does not have intersection control devices such as traffic lights. The SRN is without gated entry and exit so individual vehicles are not systematically identified as they join and leave the network. (\cite{NationalHighways2022})

The NTIS Network and Asset Model contains the information on the details and location of the different systems National Highways uses to monitor and control traffic on the SRN. It contains information on the location of MIDAS sensor sites and geospatial information of the road junctions and motorways that can be converted into a graph representation of the network. Attributes are also available to determine the direction of travel, capacity and length of the associated weighted graph's edges. (\cite{NationalHighways2022})

\begin{figure}[t]
\centering
\includegraphics[width=0.8\textwidth]{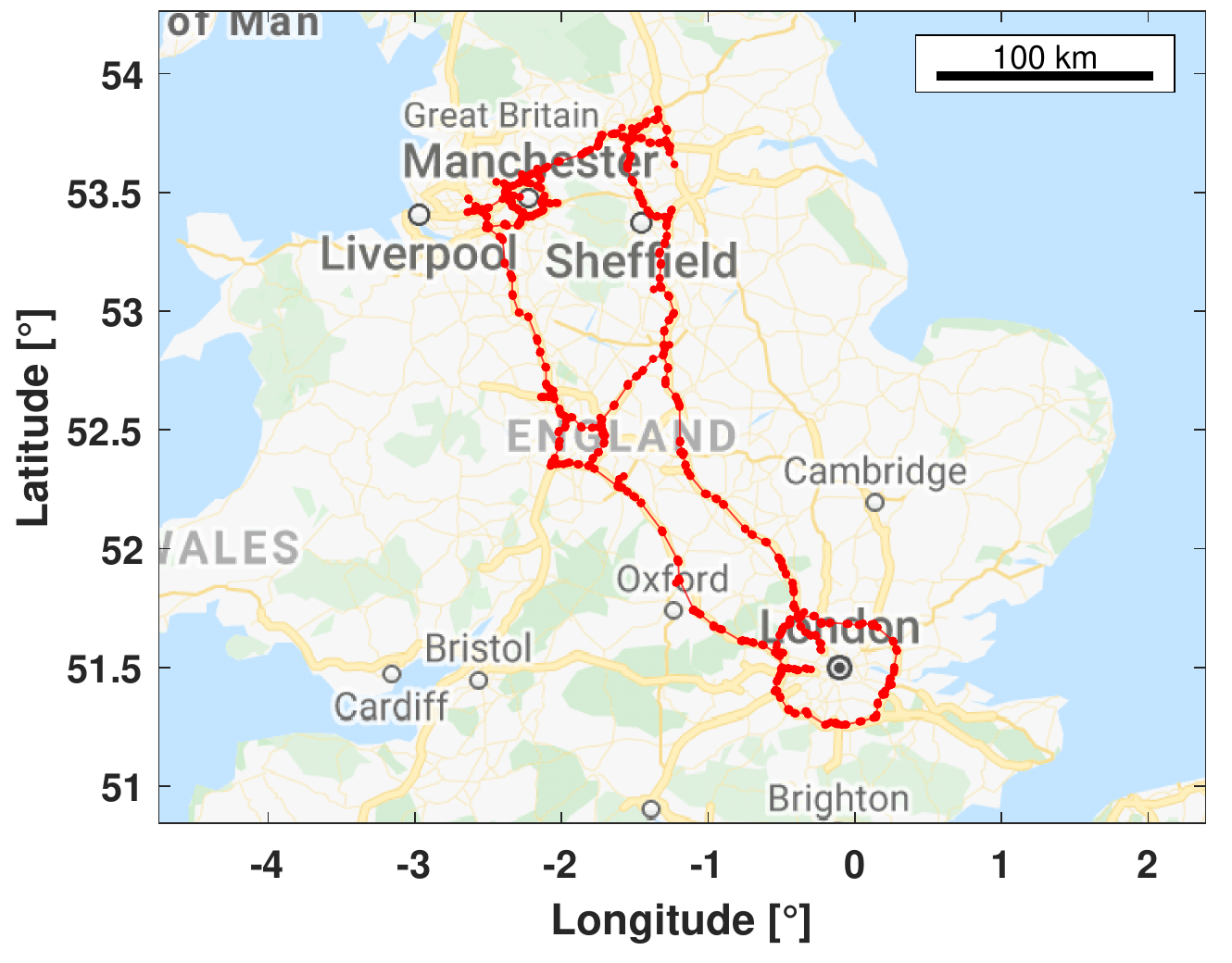}
\caption{Graph Representation of the NTIS model of the SRN in the central subnetwork area. Map underlay from Google Maps (\cite{GoogleMaps}).}
\label{fig:EnglandSRN}
\end{figure}

After conventional data selection and removal of faulty sensor observations, the central portion of the network was selected for analysis, comprising the main carriageways with relevant MIDAS sensor sites connecting a selection of major cities in England (Figure \ref{fig:EnglandSRN}). 

\begin{figure}[t]
\centering
\includegraphics[width=0.5\textwidth]{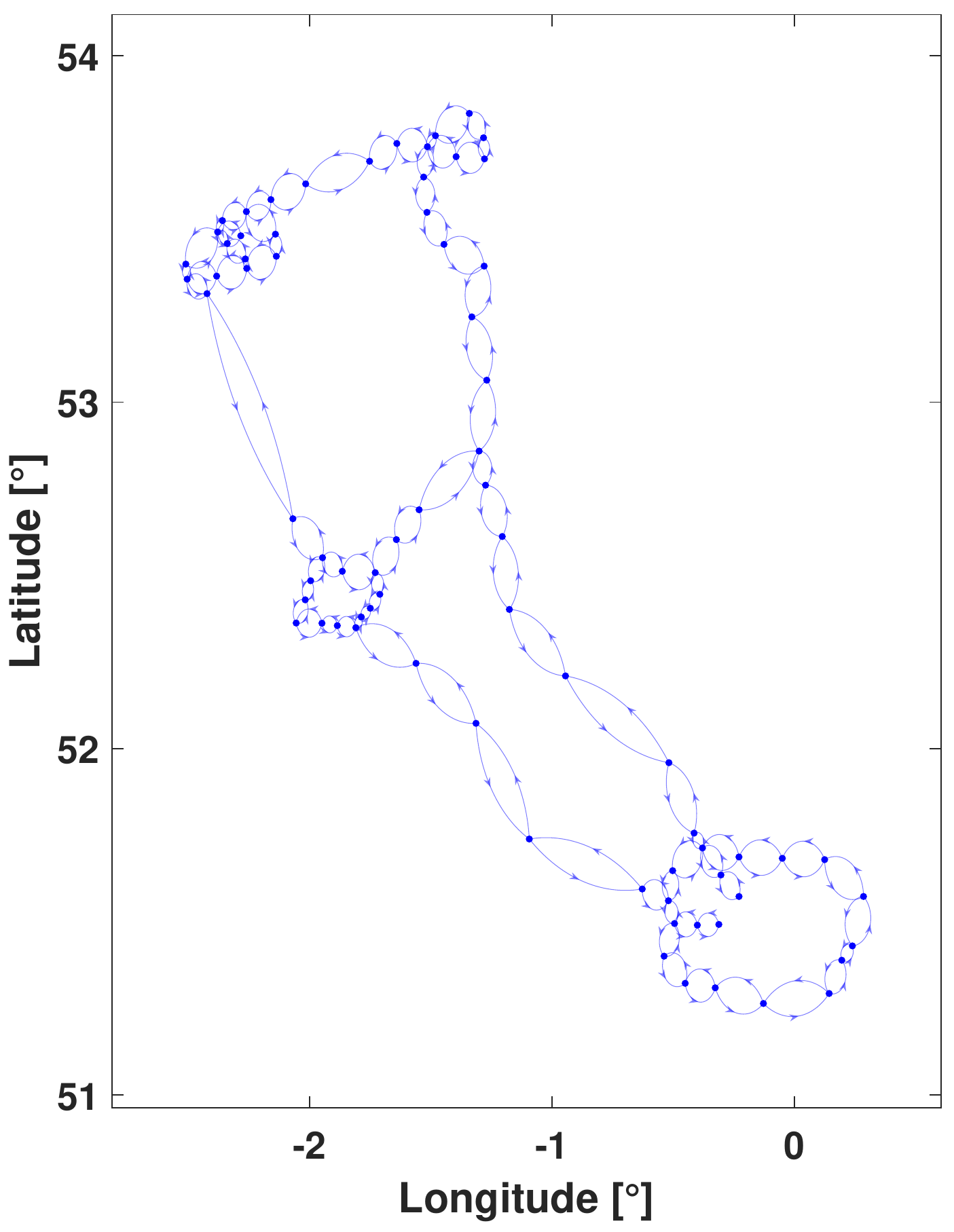}
\caption{Topographic representation for the subnetworks of the main roads connecting  the central SRN.}
\label{fig:ExpandedManBHamNetworks}
\end{figure}

\subsection{Network graph topographic representation}
The scale of the model is not concerned with navigation through the junctions between roads but instead with modelling the overall flows around the network. Therefore, an arterial road topographic representation is created for the SRN.  This creates a processed version of the NTIS model with the junctions and interchanges simplified to single “supernodes” and the carriageways in-between grouped as single “superedges” (Figure \ref{fig:ExpandedManBHamNetworks}). The use of the superedges involves averaging the flows recorded by the sensors on the edges which compose them.

\subsection{MIDAS data extraction}
MIDAS data from the available sensors is extracted and matched to the associated topographic superedge through the NTIS dataset. The flow data recorded are also grouped into bins of distinct time periods namely AM: 6am – 10am, MD (midday): 10am – 4pm, PM: 4pm – 8pm. For each time period, the mean hourly flow is calculated over the respective period.

Loop detector data can be noisy and needs to be processed correctly (\cite{Knoop2017}). When multiple sensors are available on the same superedge, the median flow readings are used. This both minimises the effect of outliers and filters out erroneous readings, as those differing from the median by more than twice the median absolute deviation. This allows the central tendency of measured flows to be resistant to sensors with faults or which do not measure the main carriageway flow even after the slipway sensors are excluded through their database names. 

The TA models are fitted to the MIDAS data taken from September 2018 to May 2019. The month of June 2019 is reserved for model validation.

%% file: Results.tex
\section{Results}

\subsection{Accuracy of different applications of the partitioning on the England SRN}

\begin{figure*}[t]
\centering
\begin{subfigure}[h]{.48\textwidth}
\includegraphics[width=\textwidth]{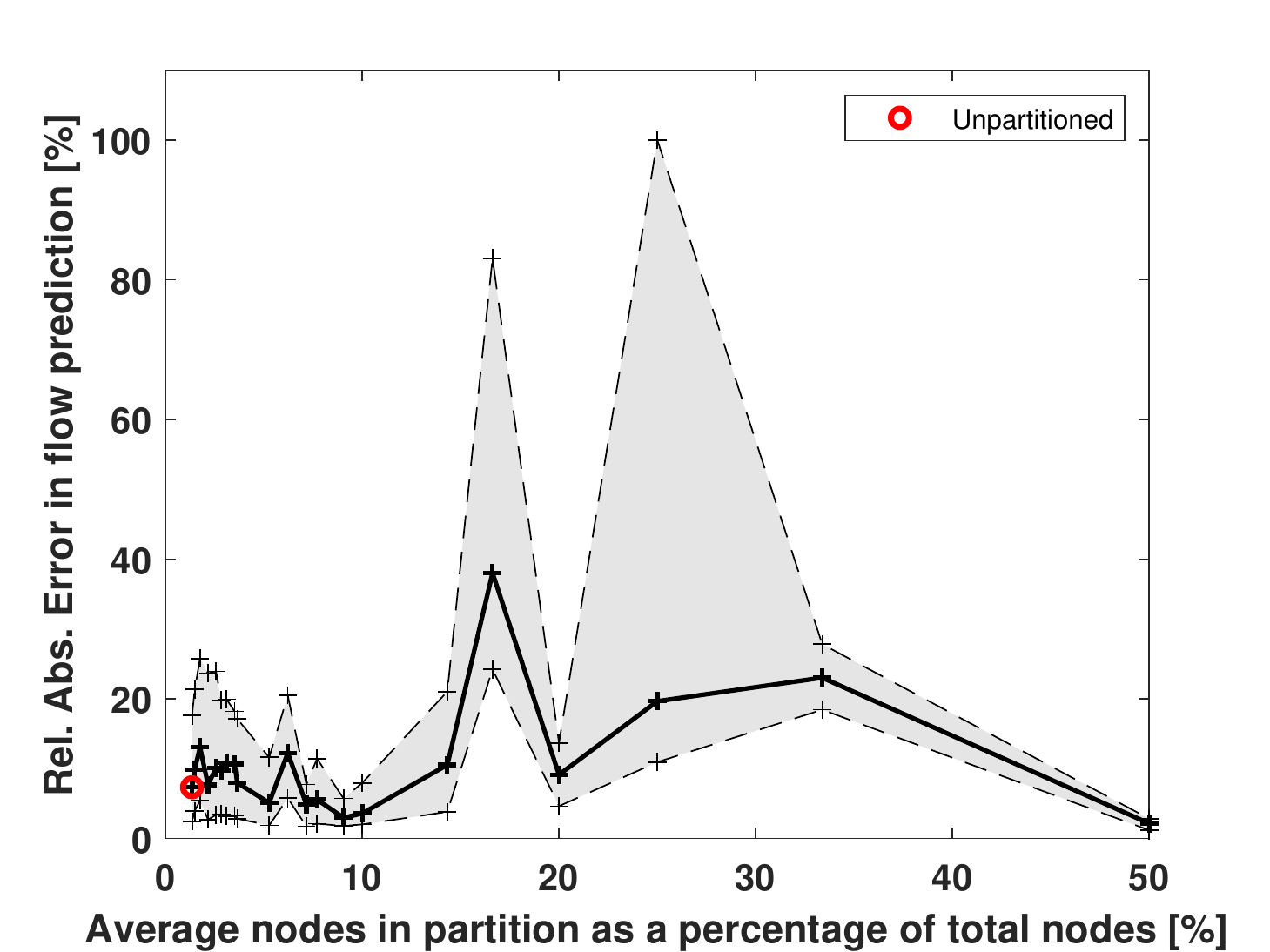}
\caption{Degenerate}
\end{subfigure}
\begin{subfigure}[h]{.48\textwidth}
\includegraphics[width=\textwidth]{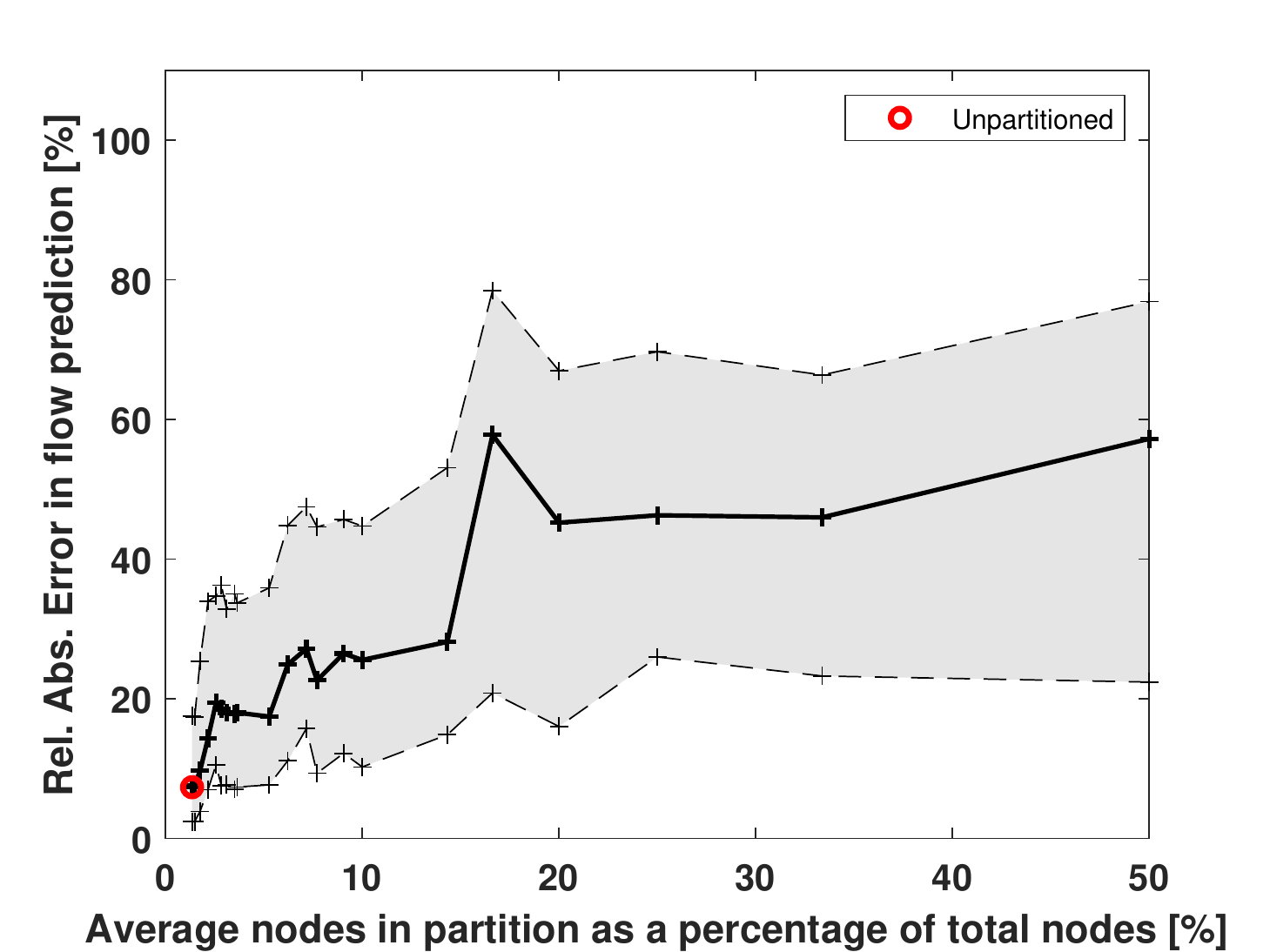}
\caption{External}
\end{subfigure}
\begin{subfigure}[h]{.48\textwidth}
\includegraphics[width=\textwidth]{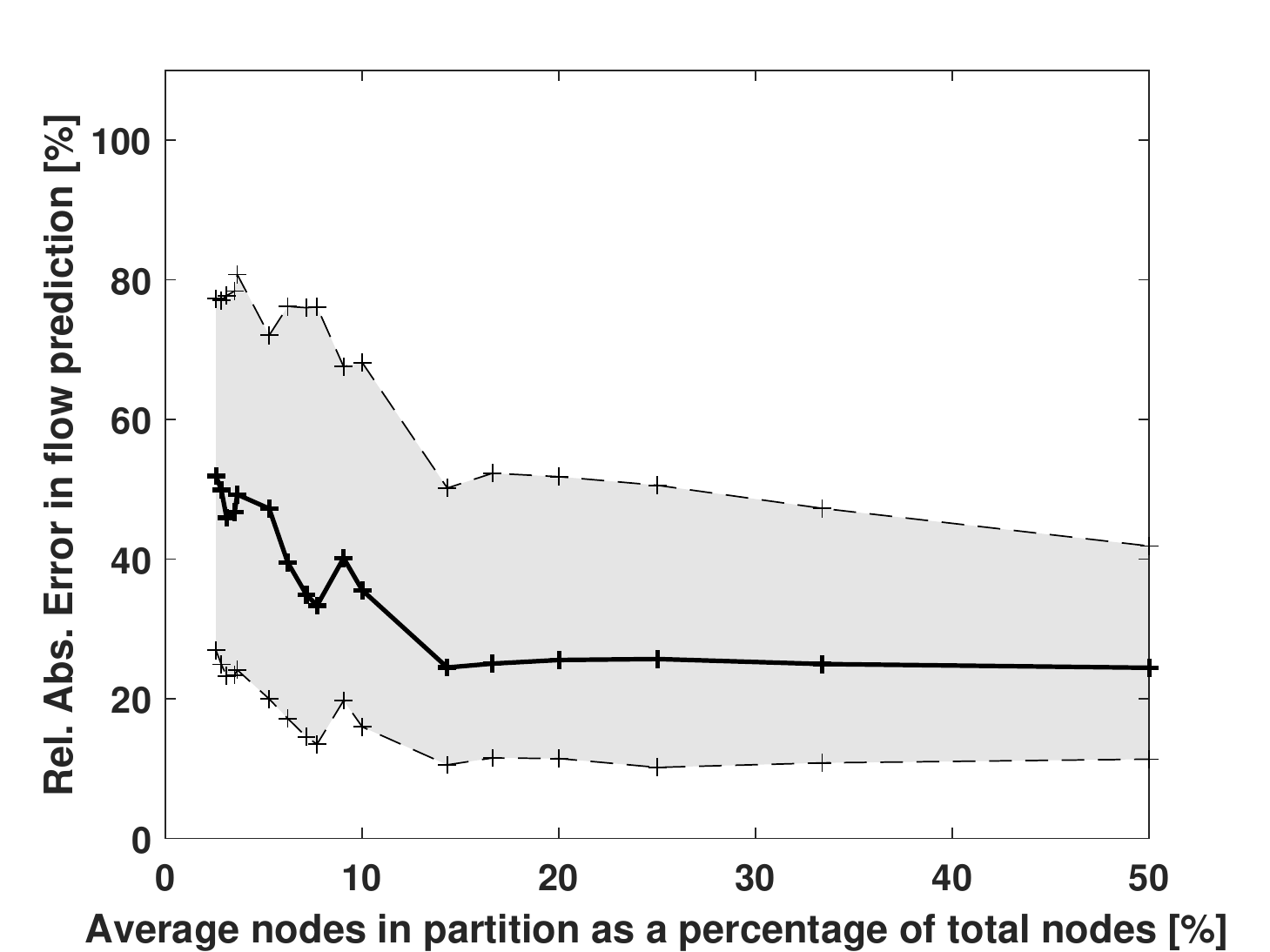}
\caption{Internal}
\end{subfigure}
\begin{subfigure}[h]{.48\textwidth}
\includegraphics[width=\textwidth]{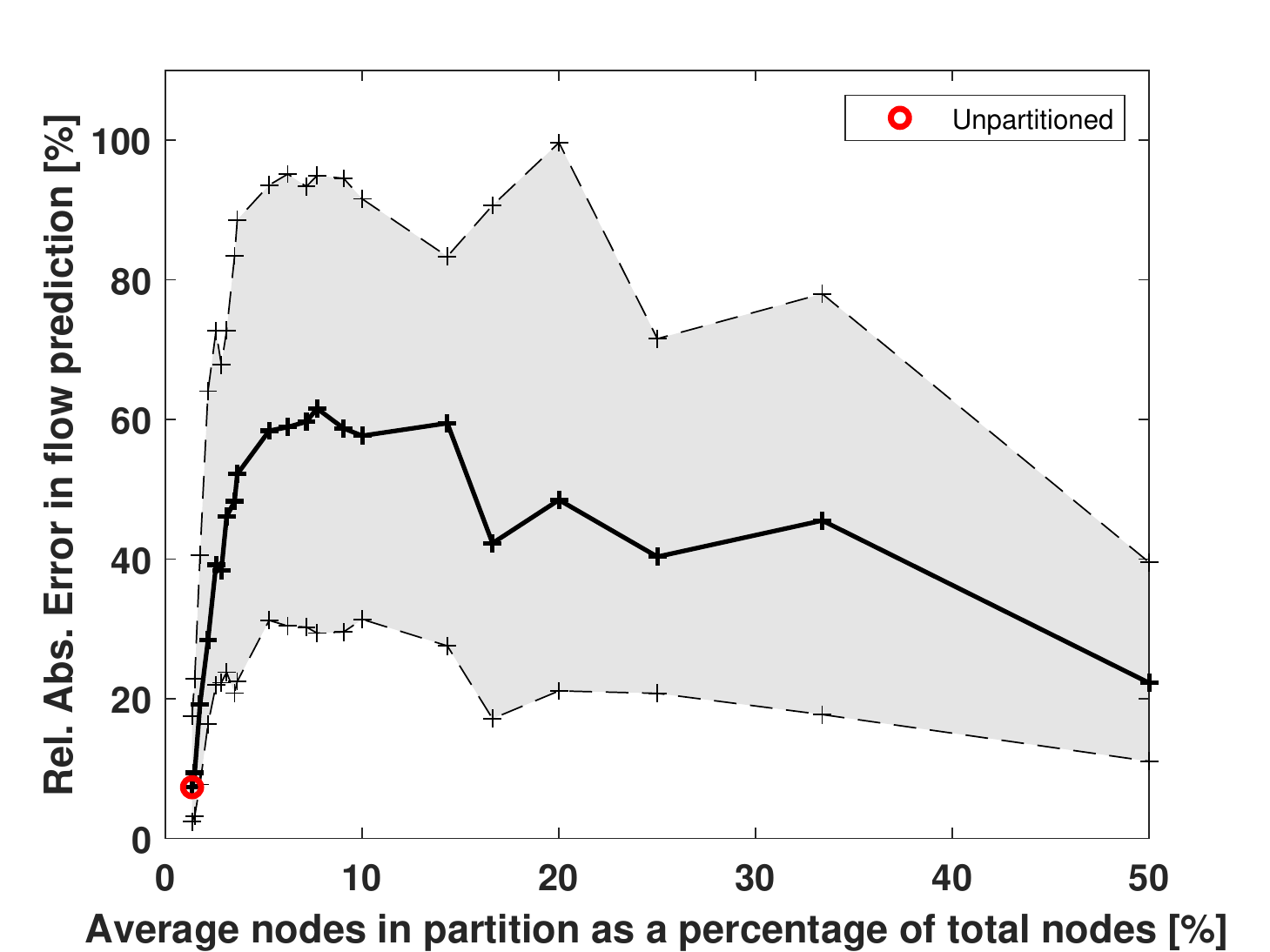}
\caption{Internal and External}
\end{subfigure}
\caption{Plot of Relative Absolute Error in user-equilibrium flow prediction for each partition size investigated on the English SRN subnetwork.  Solid line is median error and dashed lines indicate the IQR. Lines are used as visual aid for the individual point results.}
\label{fig:FlowMedianE2}
\end{figure*}

\begin{figure*}[h]
\centering
\begin{subfigure}[h]{.48\textwidth}
\includegraphics[width=\textwidth]{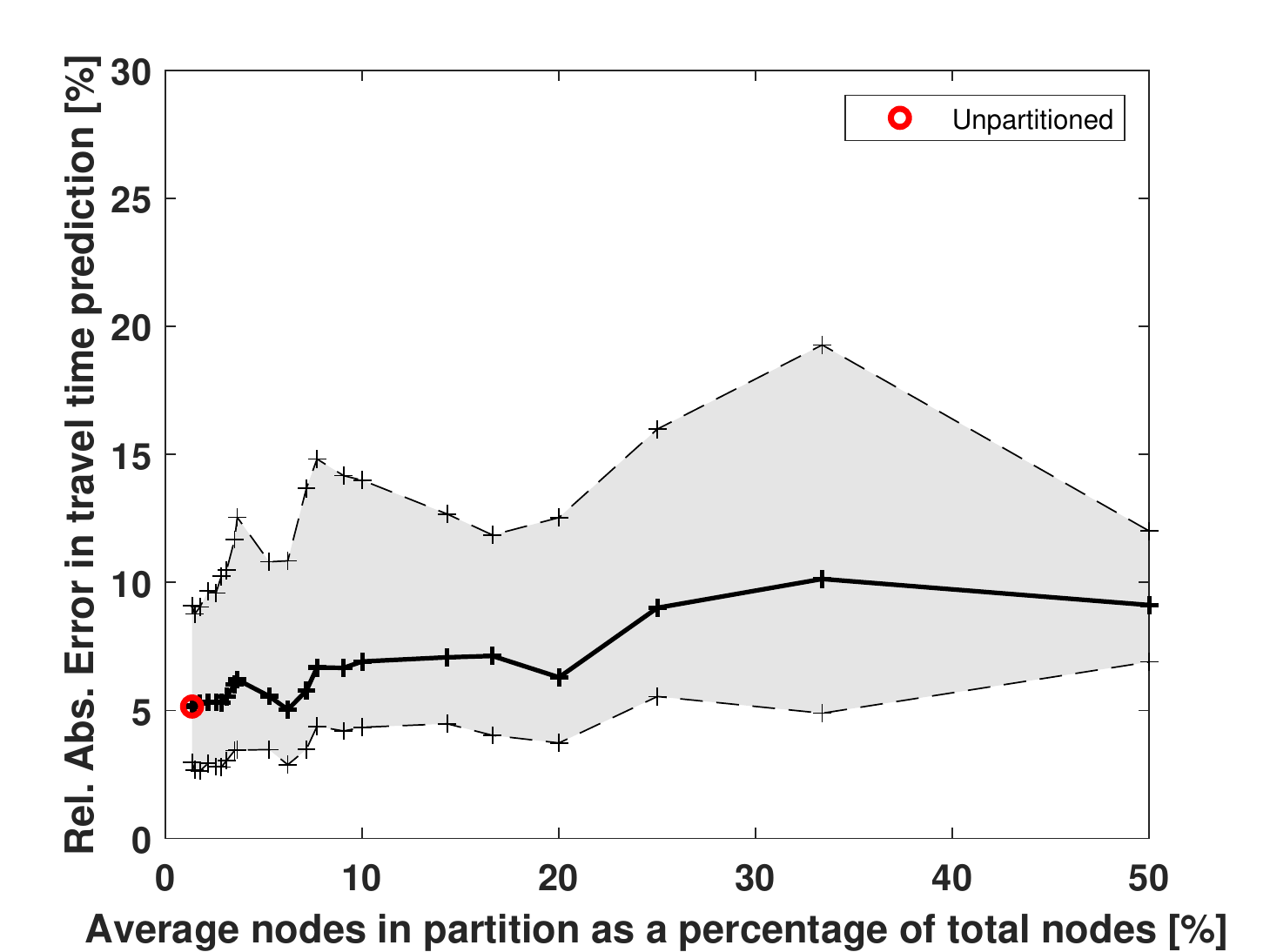}
\caption{Degenerate}
\end{subfigure}
\begin{subfigure}[h]{.48\textwidth}
\includegraphics[width=\textwidth]{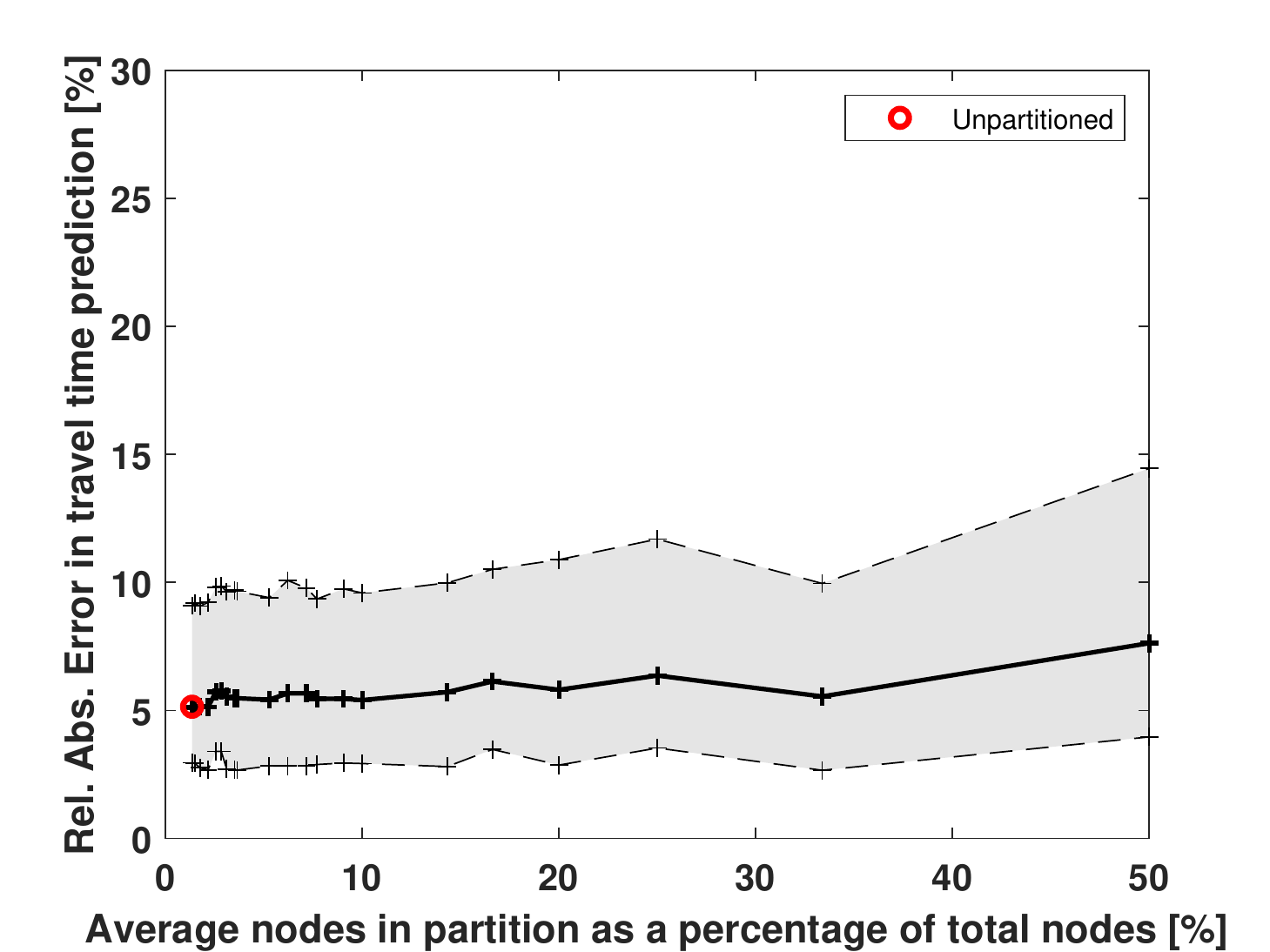}
\caption{External}
\end{subfigure}
\begin{subfigure}[h]{.48\textwidth}
\includegraphics[width=\textwidth]{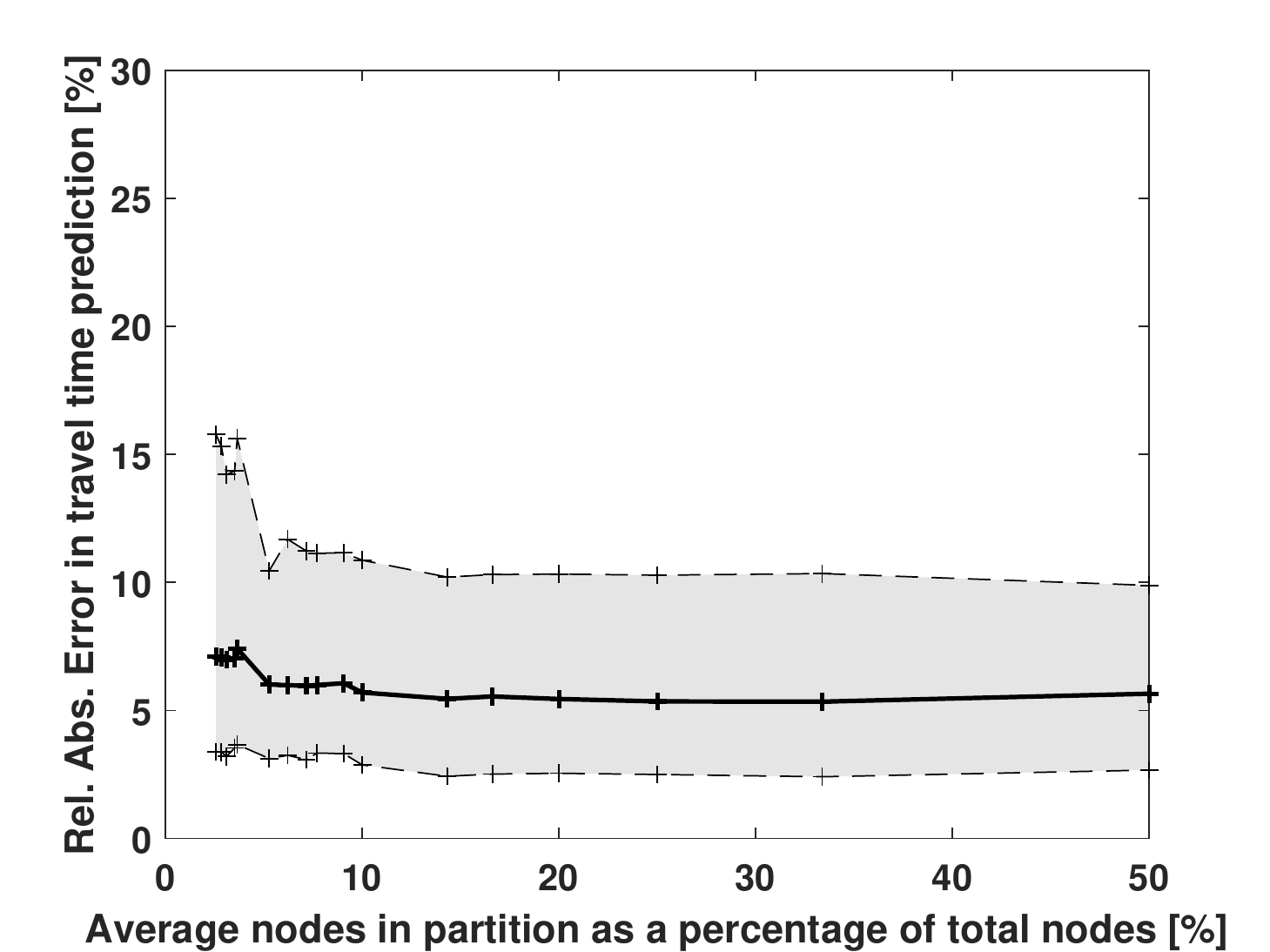}
\caption{Internal}
\end{subfigure}
\begin{subfigure}[h]{.48\textwidth}
\includegraphics[width=\textwidth]{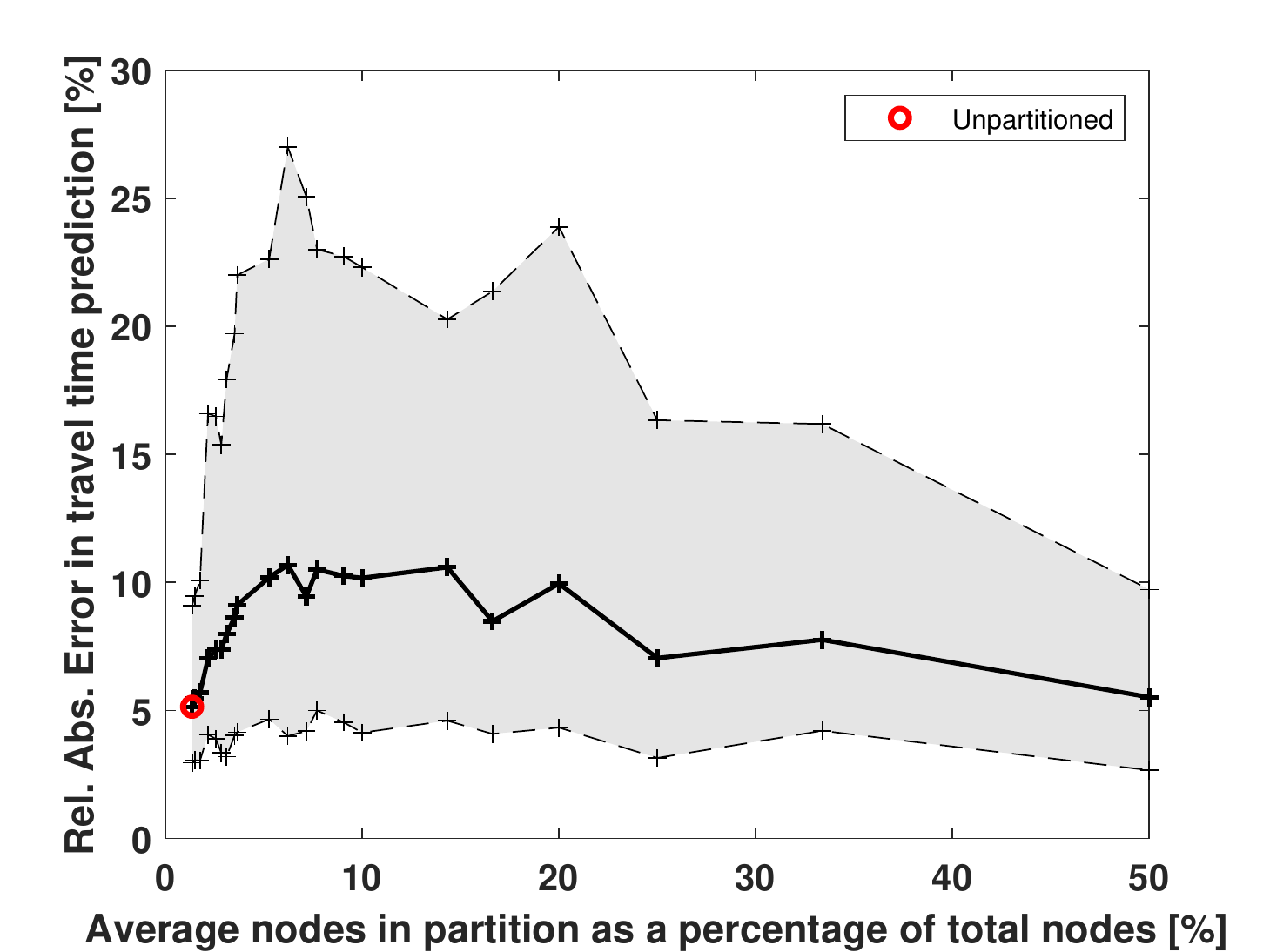}
\caption{Internal and External}
\end{subfigure}
\caption{Plot of Relative Absolute Error in user-equilibrium travel time prediction for each partition size investigated on the English SRN subnetwork. Solid line is median error and dashed lines indicate the IQR. Lines are used as visual aid for the individual point results.}
\label{fig:TimeMedianE2}
\end{figure*}

To investigate the effect of partition resolution on each of the types of partitioning matrix estimation techniques, we varied the Louvain resolution parameter and evaluated the effect on TA model accuracy and computation requirements of the resulting partitionings. The validation was performed on the reserve June 2019 month of flow and speed data for the three day time bins (AM, MD, PM). Making an assessment based on flow and travel time prediction is a practical way to validate the accuracy of the calculated O-D matrices. 

Relative errors in the flow and travel times of the user-equilibrium assignment prediction are used to evaluate the performance.  The relative absolute errors are calculated as:

\begin{equation} \label{eq:RAETime}
	RAE_a^t = \frac{\vert t_{p,a}^{user} - t_{p,a}^{obs}\vert}{t_{p,a}^{obs}}
\end{equation}\\\
for travel time, while
\begin{equation} \label{eq:RAEFlow}
	RAE_a^x = \frac{ \vert x_{p,a}^{user} - x_{p,a}^{obs}\vert}{x_{p,a}^{obs}}
\end{equation}\\\
is used for flows. For each time-bin $p$ and edge $a$, $x_{p,a}^{obs}$ is the observed flow and $t_{p,a}^{obs}$ is the travel time derived from observed speed. The values are the average within each time-bin over the 19 weekdays of the validation month. $t_{a}^{user}$ is the predicted travel time derived from the congestion function using $x_{a}^{user}$, which is the edge flow value predicted by the model through solving the TAP with the calculated O-D matrix.

The results for the four estimation approaches can be seen to exhibit different patterns as the size of the partitions varies (Figure \ref{fig:FlowMedianE2} and \ref{fig:TimeMedianE2}). The error in flow and travel time prediction can be compared to the result for the unpartitioned case which is a benchmark for the methods, giving the same value for all methods except internal-only for which it was unattainable.

Comparing the different approaches for using the partitioning, it can be seen that there is considerably different behaviour between degenerate and non-degenerate approaches (Figure \ref{fig:FlowMedianE2}). The flow prediction accuracy for degenerate varies less for the partitions with a smaller percentage of the total supernodes inside (a larger number of partitions), however as the size of the partitions increases the flow prediction has a larger variance between resolutions. The relative error for flow is low for the largest partition size. This can be attributed to the network being degenerated to a two node, two edge system so the demand prediction through GLS becomes trivial. It can be seen that the time prediction accuracy for the degenerate method deteriorates as the partitions become larger and less numerous.

\begin{figure}[t]
\centering
\includegraphics[width=.5\textwidth]{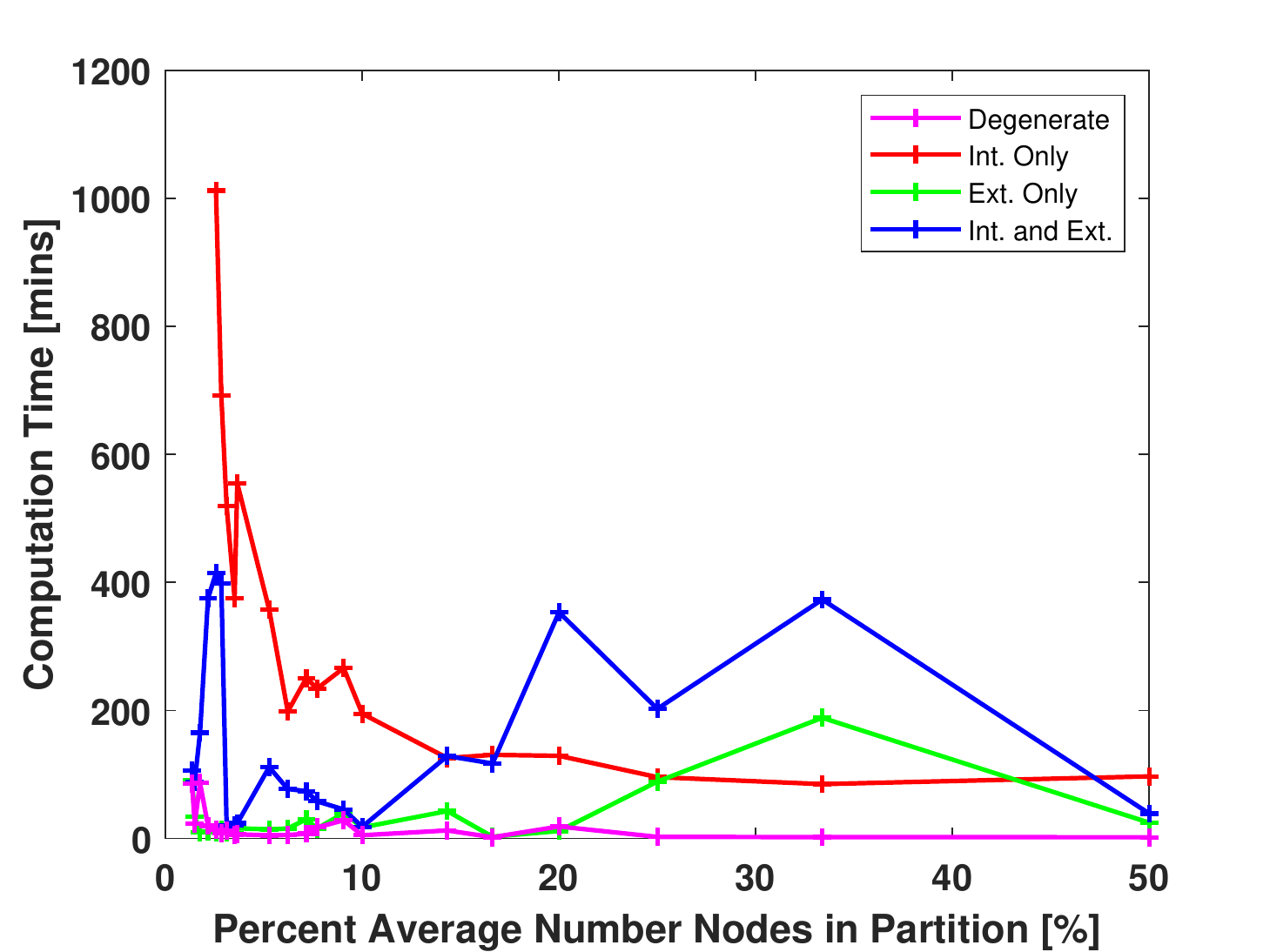}
\caption{Computation Time of TAP results for each partition size investigated on the English SRN subnetwork. Lines are used as visual aid for the individual point results.}
\label{fig:E2CompTime}
\end{figure}

Between the other non-degenerate methods (internal-only; external-only; internal-external combined) in Figure \ref{fig:FlowMedianE2} several trends can be seen. With internal-only, as the size of the partitions increases to include more supernodes the results for both flow and time improve up to the 15\% point. Between 15-50\% the median is approximately constant. In Figure \ref{fig:E2CompTime} the computation time for internal-only also levels off past the 15\% point. This implies the results for using the internal-only approach are similar for the 15-50\% partition size range in both accuracy and computation time. The results for internal-only were not available for the smallest five resolutions of partitioning. This is because the estimate of the prior matrix was too inaccurate for the O-D adjustment algorithm to converge.

As the percent of average supernodes in a partition increases, the results for the external-only method show a broadly linear increase in error for flow and time prediction as the computation time stays mostly low. This is due to the prior matrix increasingly basing the individual O-D movements on a smaller subset of topographic superedges. Less information is available so the prior matrix moves further from its best estimate which is the unpartitioned case.

When internal and external estimates are combined to create the prior matrix, it can be seen that there is a degradation in accuracy for flow and time prediction from the unpartitioned case to approximately the 7\% of total supernodes point. After this, the results for both flow and time improve almost linearly with increases in the partition size. At the largest partition size it can be seen that the accuracy matches the internal-only result but with less computation time.

For the road subnetwork, the memory requirement of the four techniques for all partition sizes did not vary much, staying being between 21.4-21.5GB in all cases. The road network is not particularly large (74 nodes) so memory is not the concern. The calculations for the results were all performed on a Dell PowerEdge C6320 with 2.4GHz Intel Xeon E5-2630 v3 CPU. The implication of the results is that the best option would be to use the largest partition possible with the internal-external combined or the internal-only methods. 

\subsection{Comparison of the results with different sized theoretical networks}
To investigate how the size of the network influences the results of the different methods the same tests were carried out on additional theoretical networks of a range of sizes (see appendix for details) and the Sioux Falls test network commonly used in TA model testing (\cite{TransportationNetworksGithub}). The validation was carried out on simulated flow data without travel time.

\begin{figure*}[h]
\centering
\begin{subfigure}[h]{.48\textwidth}
\includegraphics[width=\textwidth]{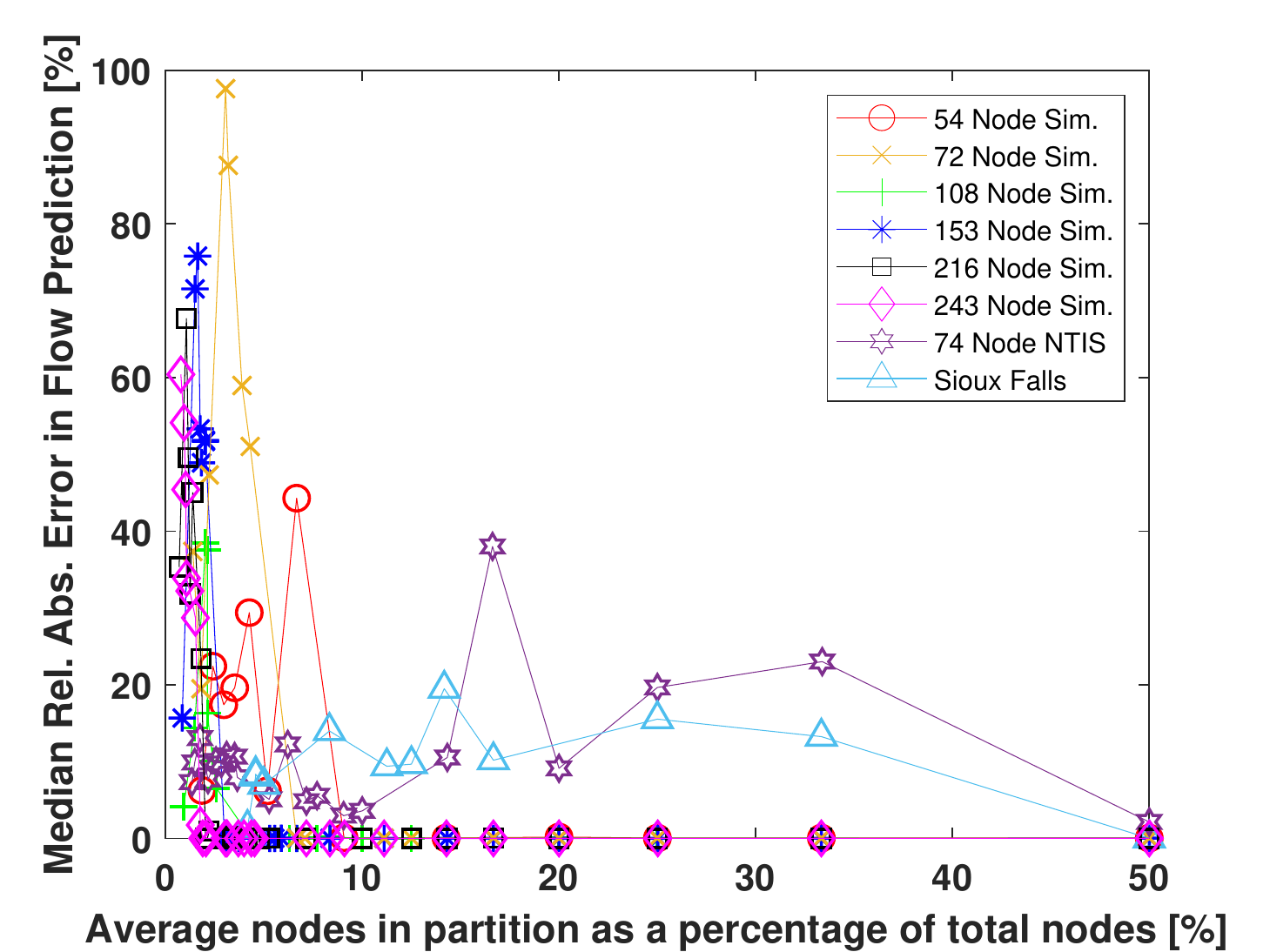}
\caption{Degenerate}
\end{subfigure}
\begin{subfigure}[h]{.48\textwidth}
\includegraphics[width=\textwidth]{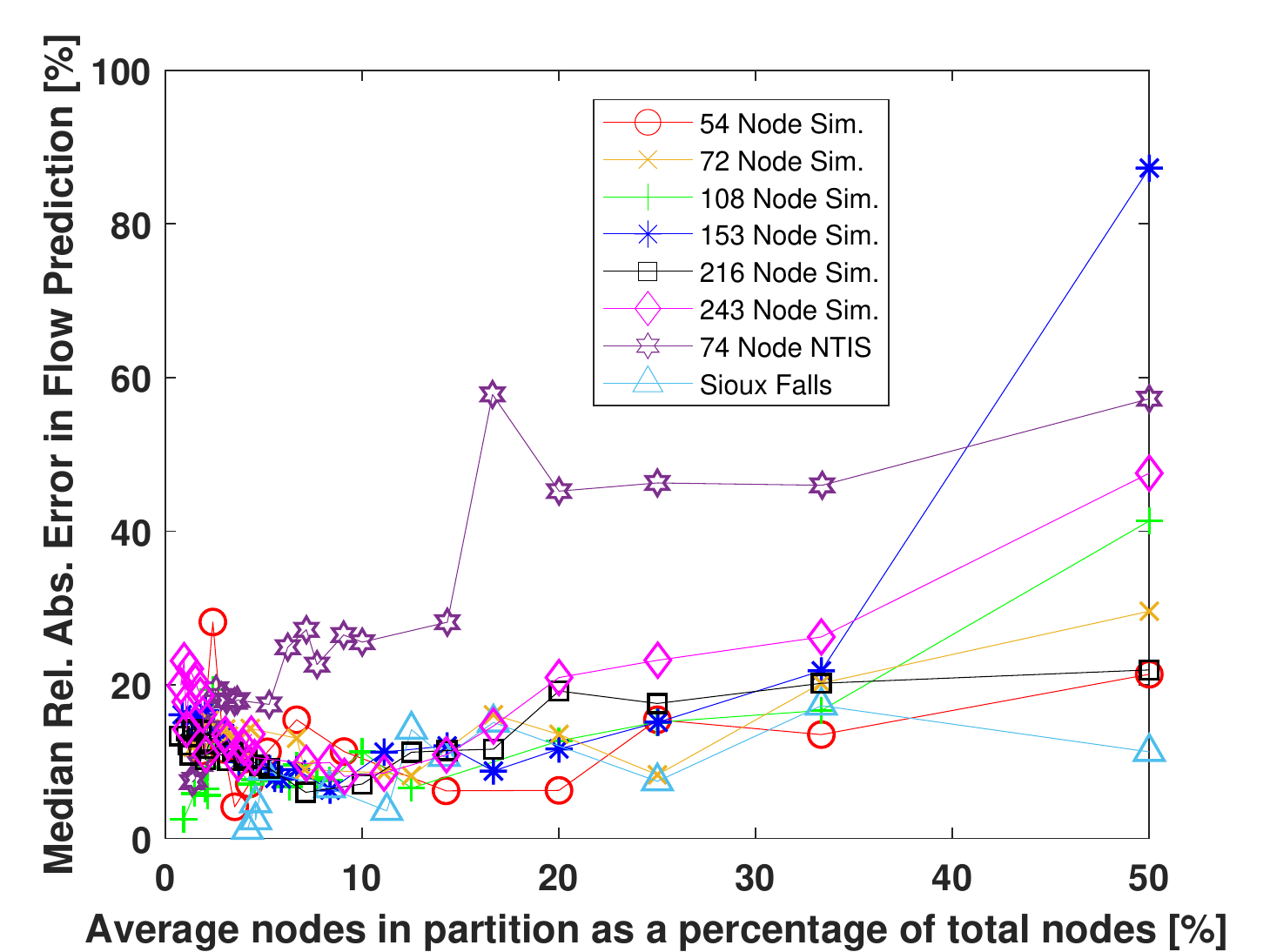}
\caption{External}
\end{subfigure}
\begin{subfigure}[h]{.48\textwidth}
\includegraphics[width=\textwidth]{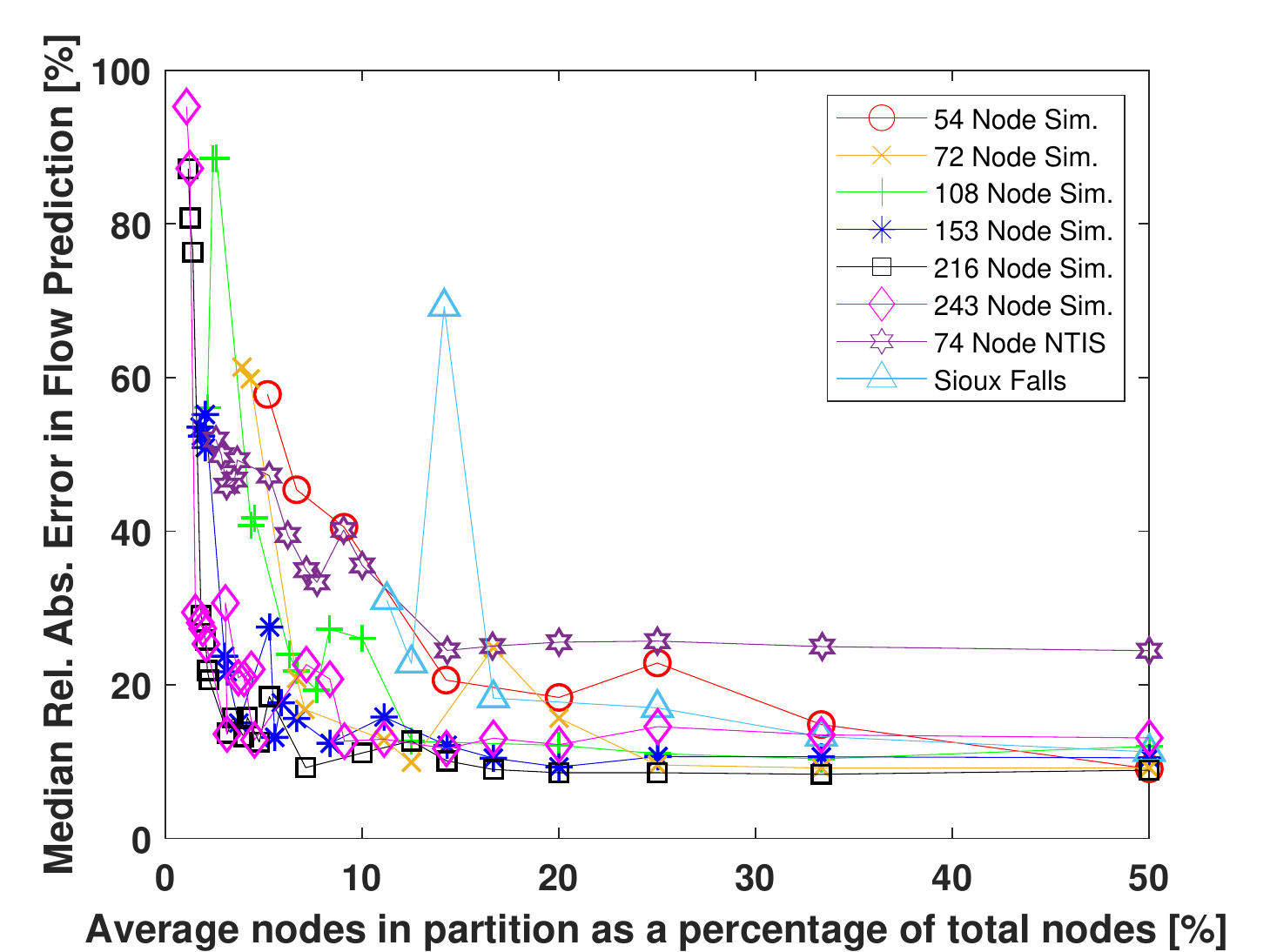}
\caption{Internal}
\end{subfigure}
\begin{subfigure}[h]{.48\textwidth}
\includegraphics[width=\textwidth]{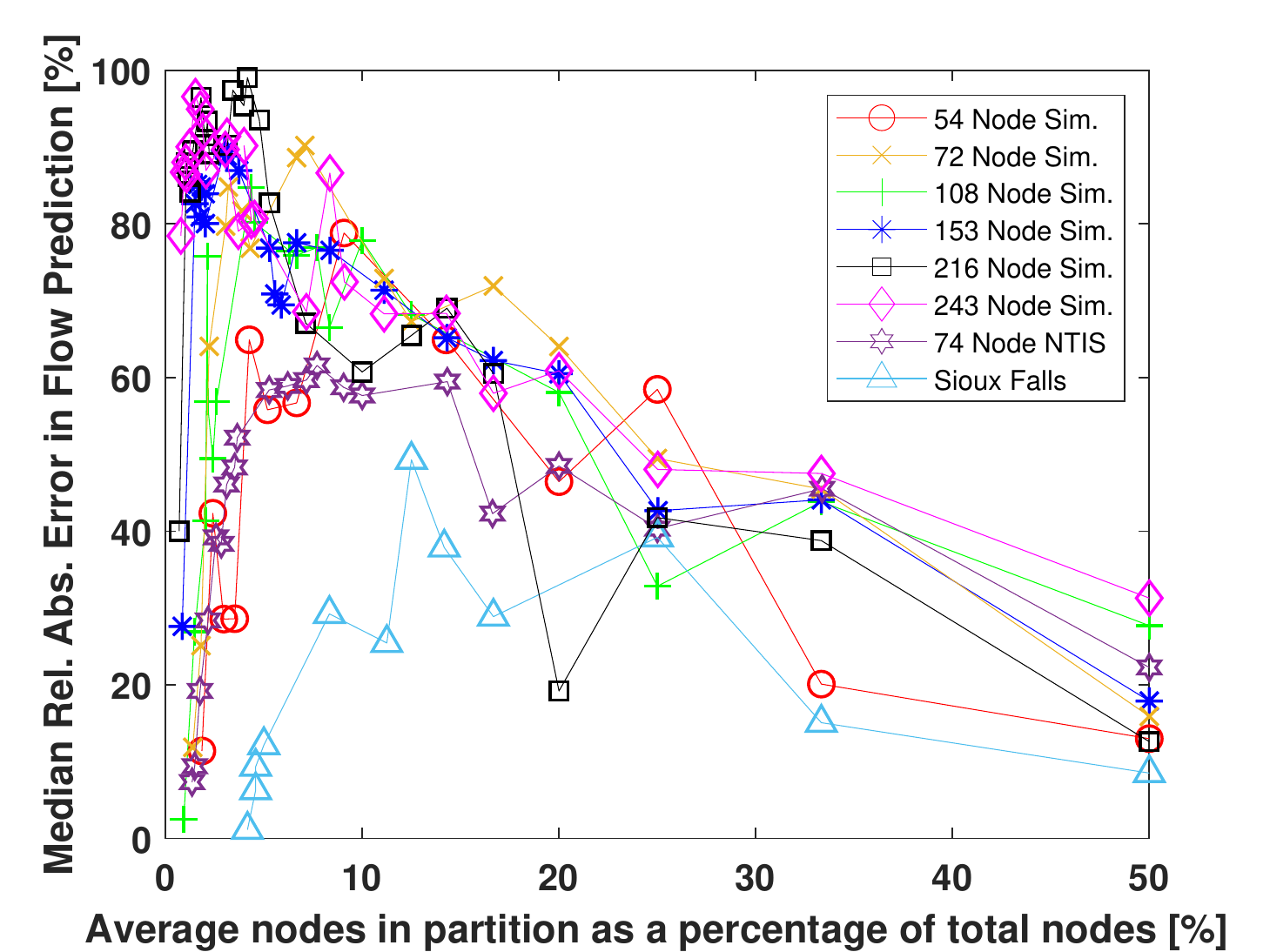}
\caption{Internal and External}
\end{subfigure}
\caption{Plot of Median Relative Absolute Error in user-equilibrium flow prediction for each partition size investigated on the theoretical networks. Lines are used as visual aid for the individual point results.}
\label{fig:FlowMedianMN}
\end{figure*}

Similar trends to the English SRN network application can be seen when they are applied to the theoretical networks (Figure \ref{fig:FlowMedianMN}). For internal-only, there is a peak in error for small partition sizes with no results produced for the smallest partitions. The internal-only results level out after 15\%. For external-only there is a steady increase in flow error as the partition size increases. The results for the internal-external combined method show the same characteristic triangle shape with an initial increase followed by a decrease in error.

For the degenerate method the trend is similar for Sioux Falls but different for the theoretical networks. With the theoretical networks, there is a peak in error between 0-10\% and then the error reduces to almost zero for the larger partition sizes. We can attribute this to the theoretical networks having no congestion and the simulated flows being created with a Poisson distribution so that for the smaller network sizes (larger partitions) very accurate estimates of the demand are obtained.

\subsection{Computational requirements}
We investigated the computational requirements of our partitioning approaches using the theoretical networks.
\subsubsection{Computational requirements without partitioning}
\begin{figure*}[h]
\centering
\begin{subfigure}[h]{.48\textwidth}
\includegraphics[width=\textwidth]{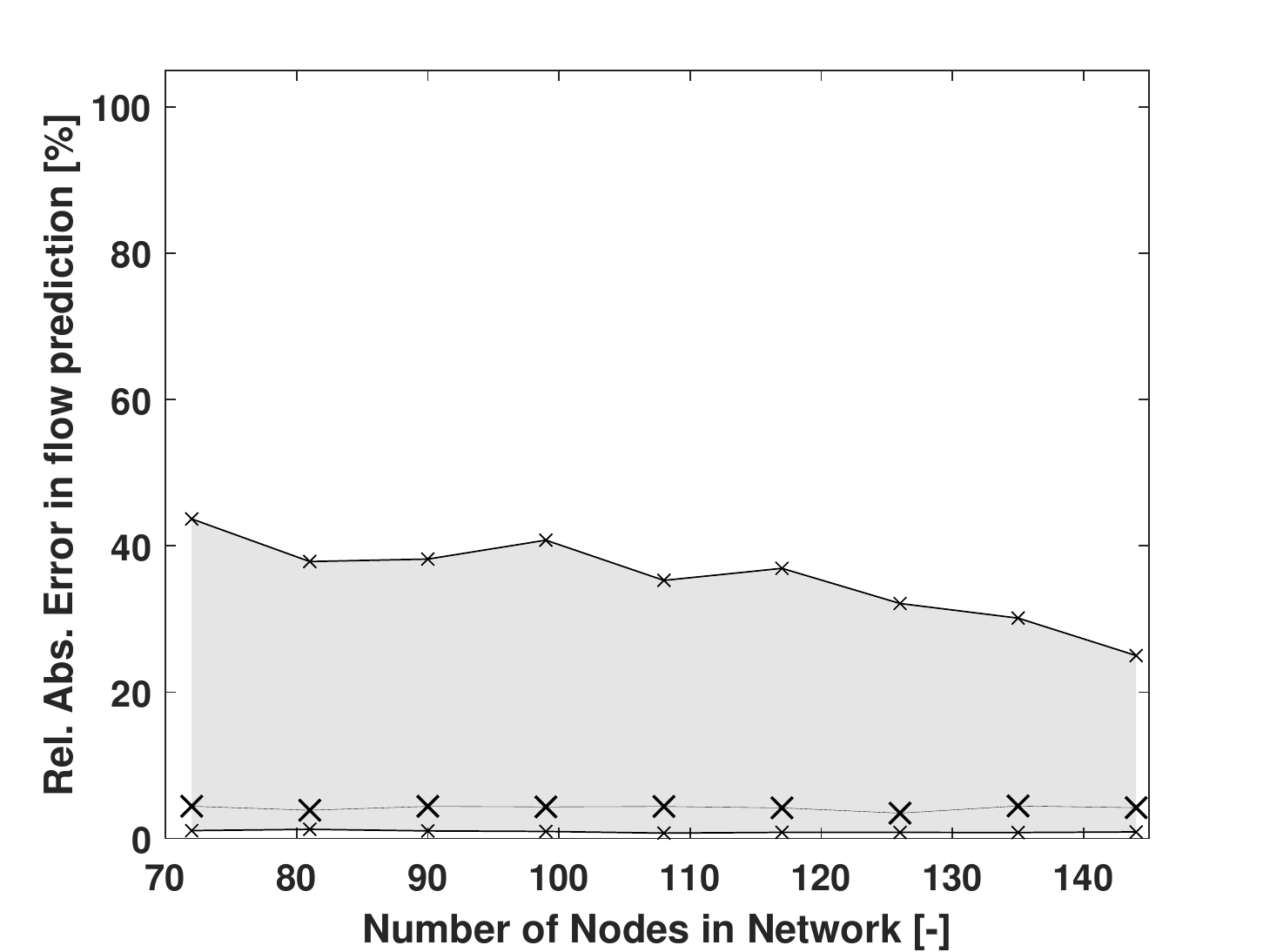}
\caption{User Equilibrium Flow Prediction}
\end{subfigure}
\begin{subfigure}[h]{.48\textwidth}
\includegraphics[width=\textwidth]{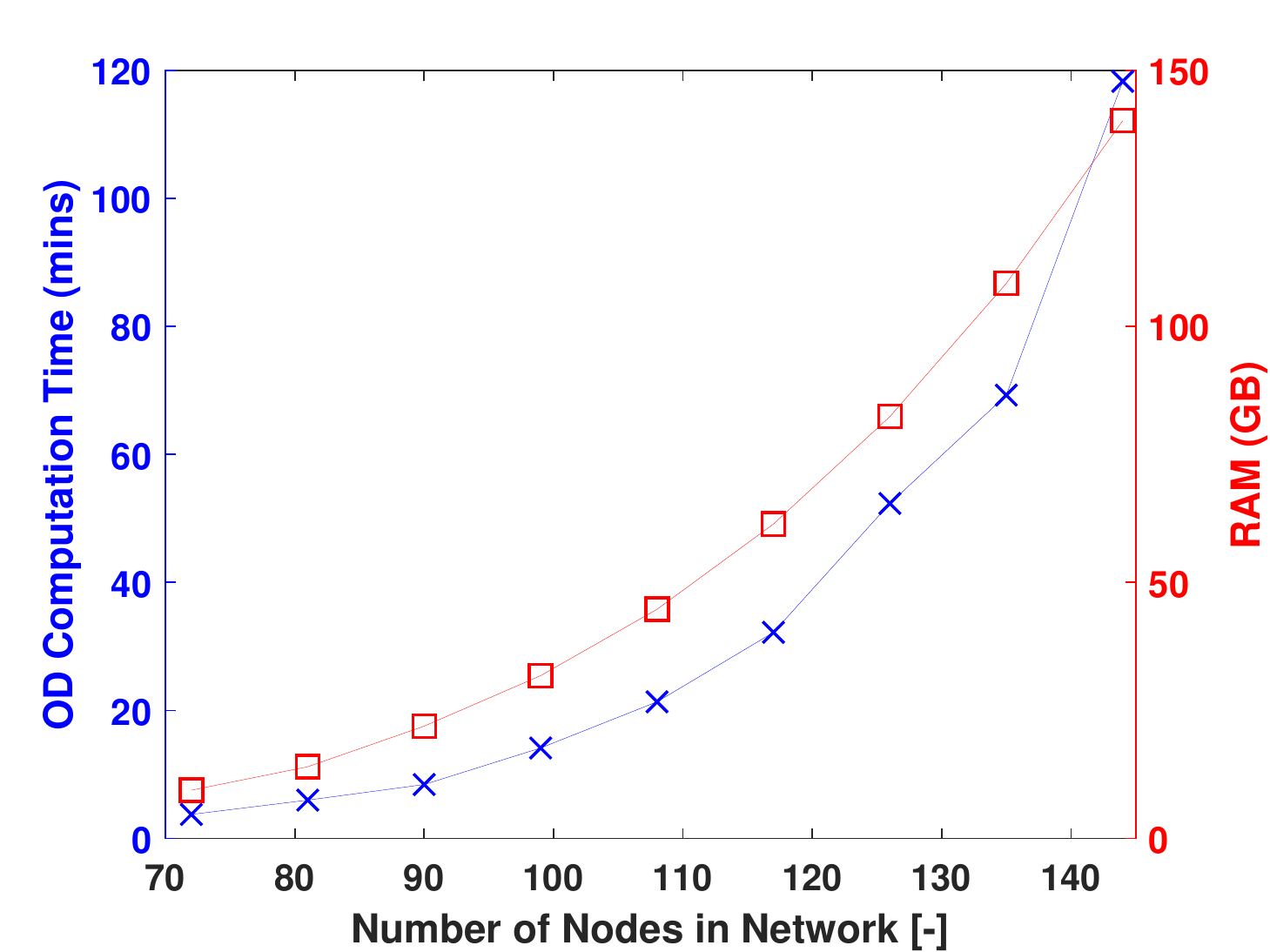}

\caption{Computational Requirements}
\end{subfigure}
\caption{Flow prediction error and memory requirements for a range of network sizes when the O-D estimation and adjustment are applied to a range of networks without the use of partitioning. In (a) the solid line is median error and dashed lines indicate the IQR. Lines are used as visual aid for the individual point results.}
\label{fig:unpart_networks}
\end{figure*}

When the GLS method of O-D estimation is applied to a network without partitioning being used it can be seen in Figure \ref{fig:unpart_networks} that the median error in flow prediction remains constant as the network size grows but the required computational time and memory increases steeply. For the results in Figure \ref{fig:unpart_networks}, no partitioning was applied at any time with the O-D estimation and adjustment algorithms being applied to the entire network. Due to the steeply increasing computational requirements, there is a limit on the number of nodes which O-D estimation can be applied to at one time.

\subsubsection{Computational requirements on larger networks with partitioning}
\begin{figure*}[h]
\centering
\begin{subfigure}[h]{.48\textwidth}
\includegraphics[width=\textwidth]{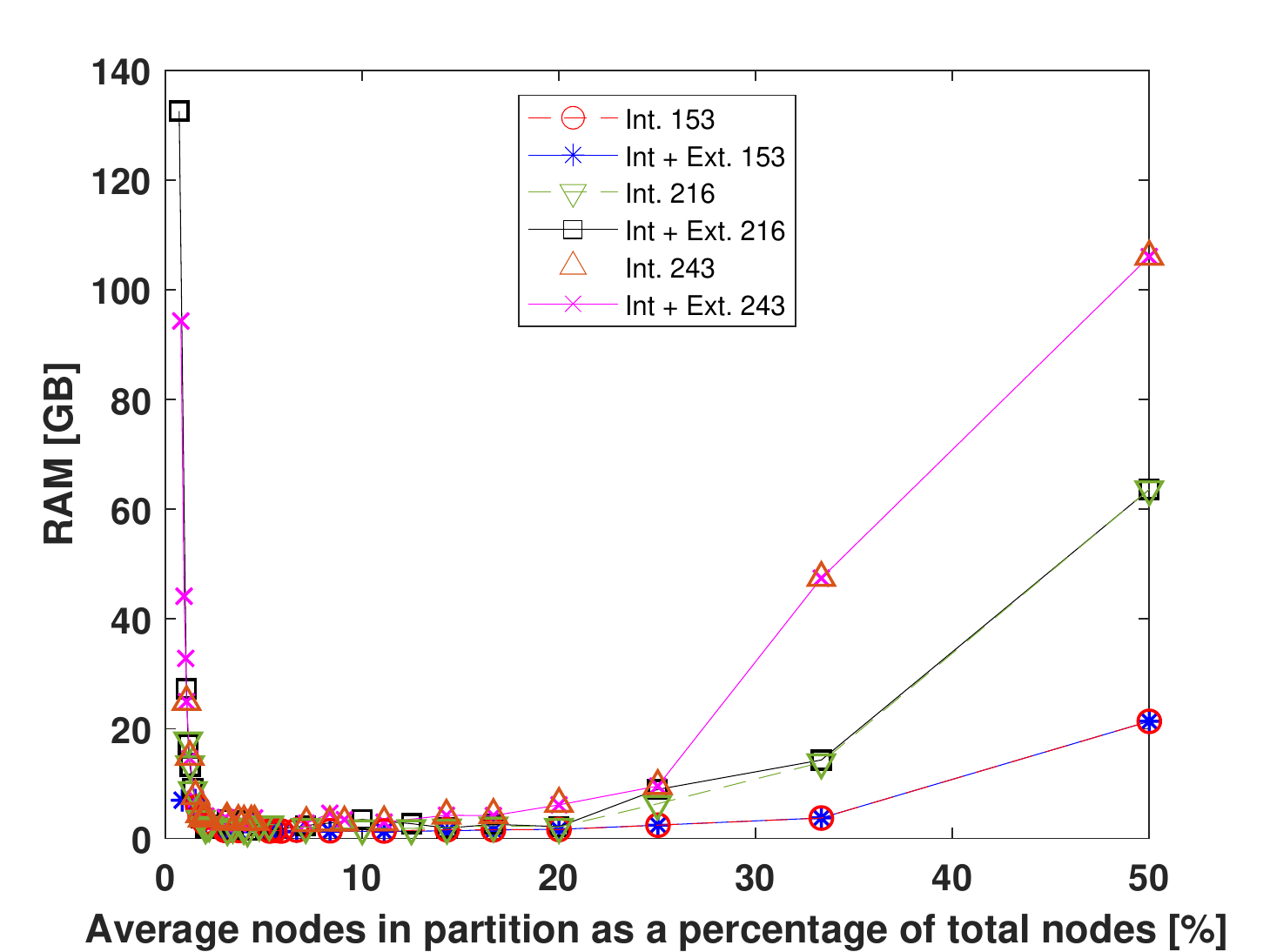}
\caption{Memory}
\end{subfigure}
\begin{subfigure}[h]{.48\textwidth}
\includegraphics[width=\textwidth]{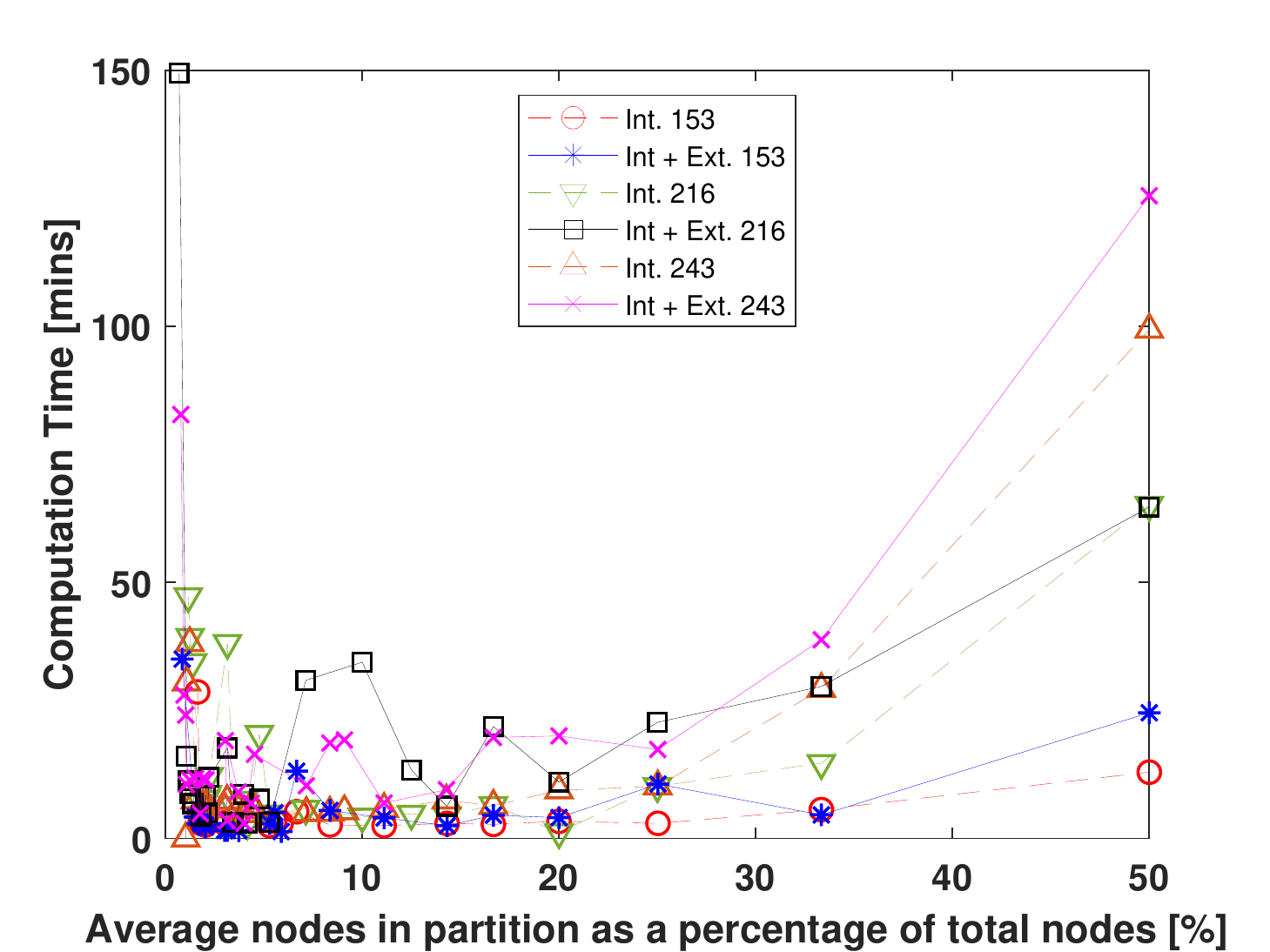}
\caption{Computation Time}
\end{subfigure}
\caption{Computational requirements for each partition size investigated for a 153, 216 and 243 node theoretical network. Lines are used as visual aid for the individual point results.}
\label{fig:RAMMN216}
\end{figure*}

We expanded the analysis of the simulated networks to larger sizes for the internal-only and internal-external combined methods which are the best performers of the non-degenerate partitioning approaches. As the networks grow in size it can be seen in Figure \ref{fig:RAMMN216} (a) that the memory requirements for both of the methods increases at the extreme ranges of partitioning. Comparing between Figures \ref{fig:unpart_networks} and \ref{fig:RAMMN216} the effectiveness of using partitioning to reduce the computational requirements for larger networks can be seen. For example, by using partitioning (internal-only and internal-external) the 243 network uses a similar amount of RAM and computational time to the unpartitioned 144 network.

At very small partitions the memory requirements increase very steeply. The 216 and 243 node networks were unable to be calculated unpartitioned, this is due to the size of memory required and limitations with the Gurobi solver used. Of most interest is the increase in memory at the largest partition sizes. It can be seen that as the total network size grows the memory for the larger partitions starts to become very high as each subnetwork within a partition is larger. 

This has the implication that for larger networks it would be best to choose smaller and more numerous partitions. The optimal size and number of partitions depends on the size of the overall network. Computation time (Figure \ref{fig:RAMMN216} (b)) showed a similar trend to memory for the two methods.

%% file: Conclusion1.tex
\section{Discussion}

In proposing a method of partitioning a road system through network modularity, in this work we have demonstrated its potential for the calculation of the key O-D demand input for static TA models from loop detector data. This opens up the opportunity to estimate flow patterns for large national highways systems without the need for other data sources.

The results show that partitioning the network into small communities of nodes is tolerable for a degenerate approach to reduce the size of the network being analysed. This degenerate approach could be well suited for use in infrastructure assessment models such as NISMOD where the scale of analysis is more coarse, for instance at the inter-city level.

Applying partitioning in a non-degenerate way showed that a similar level of error in user-equilibrium flow and travel time predictions can be obtained by dividing the network into a small number of larger partitions. The results show that the best accuracy results came from only using the internal O-D estimates of the partitions for the larger partition sizes. However, the results show that by also including the external partition estimates there can be a reduction in computation time in some cases. For the English SRN case study it appears that the best option is to partition the network into two large communities. In very large networks where the size is such that a two community partition is still infeasibly large, the results show that for community partitions numbering three and greater it would be better to use the internal-only approach unless the communities contain such a small proportion of the nodes that the flow error starts to rise (approx. 12.5\% of nodes or eight community partitions).

The non-degenerate approach is useful for application in more detailed traffic planning. The traffic assignment models which it can create are well suited to estimating alternative flow patterns of vehicles such as system-optimal, under which the global travel cost of all drivers is minimised through the routes they are assigned. This can be used for producing performance comparisons of different national road systems through metrics such as the Price of Anarchy and evaluating network improvement options (\cite{Youn2008}). 

The performance of the methods in this work is assessed by the prediction accuracy of the TA models using the estimated O-D matrices. The O-D matrices produced are not necessarily close representations of the true demand profile. The matrix obtained through the partitioning provides the prior matrix for the O-D adjustment algorithm to create a suitable demand input for the TA model to predict flows and travel times with the accuracy presented.

Future work could look to apply this type of multi-scale demand estimation with alternative techniques to GLS, which may be more suitable. Further research could look into how to incorporate separate terms in the O-D adjustment for the internal and external estimations of the prior matrix. Future work could also incorporate other data sources to inform the division between the O-D pairs of the externally estimated O-D movements. For example, in the AM period a greater share of demand could be distributed to the destinations where more employment is located. In this work, we used the standard formulation and coefficients for the congestion functions, more accuracy is possible through the use of more advanced function estimates.

\section{Conclusion}

In this work we developed a method of network partitioning through modularity to estimate O-D demand matrices for large road networks to be used in static TA models. We applied it to the central subnetwork of the English SRN and several theoretical networks to allow different levels of partition resolution to be tested for their effects on the results of TA models derived solely from loop detector traffic data.

We show that the approach developed allows for traffic to be analysed nationally at different scales. It can be used within infrastructure models to improve their analysis of congestion. It can also be used to create static traffic assignment models for strategic analysis and planning with a data source accessible to many transport planners. Future investigations could seek to implement the technique with more accurate techniques for O-D estimation from link counts and improved adjustment algorithms.

\section{Acknowledgements}
The authors would like to express their thanks to the University of Sheffield and the Agency for Science, Technology \& Research (A*STAR) under the A*STAR Research Attachment Programme (ARAP) for funding this project. The authors are also thankful to National Highways (England) and MWayComms for assistance providing the raw data used in the article.
\section{Author Contributions}
The authors confirm contribution to the paper as follows: study conception and design: A. Roocroft, M.A. Bin Ramli, G. Punzo; data collection and coding: A. Roocroft; analysis and interpretation of results: A. Roocroft, M.A. Bin Ramli, G. Punzo; draft manuscript preparation: A. Roocroft, M.A. Bin Ramli, G. Punzo. All authors reviewed the results and approved the final version of the manuscript.
\section{Data Availability}
Data in this study have been provided by National Highways (England) via a data sharing agreement that does not allow further distribution of the data. Requests for data should be made to National Highways to whom the MIDAS and NTIS datasets used in this work belong. The output from the models of this work can be provided upon request.
\section{Compliance With Ethical Standards}
\textbf{Conflict of interest}: On behalf of all authors, the corresponding author states that there is no conflict of interest.

%% file: Appendix.tex
\section*{Appendix}\label{secA1}
\subsection*{Creating the theoretical networks}

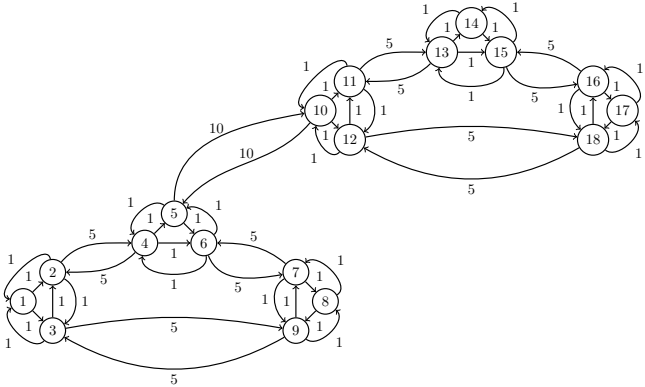
\begin{figure}[H]
\centering
\scalebox{0.55}{\begin{tikzpicture}[node distance={10mm}, thick, main/.style = {draw, circle}] 
\node[main] (1) {$1$}; 
\node[main] (2) [above right of=1] {$2$};
\node[main] (3) [below right of=1] {$3$}; 
\node[main] (4) [above right of=2, xshift=1.5cm] {$4$};
\node[main] (5) [above right of=4] {$5$}; 
\node[main] (6) [below right of=5] {$6$};
\node[main] (7) [below right of=6, xshift=1.5cm] {$7$};
\node[main] (8) [below right of=7] {$8$}; 
\node[main] (9) [below left of=8] {$9$};

\draw [->] (3) -- node[midway, right] {1} (2);
\draw [->] (2) to [out=330,in=30] node[midway, right] {1} (3);
\draw [->] (1) -- node[midway, below left] {1} (3);
\draw [->] (3) to [out=230,in=210] node[midway, below left] {1} (1);
\draw [->] (1) -- node[midway, above left] {1} (2);
\draw [->] (2) to [out=100,in=170] node[midway, above left] {1} (1);

\draw [->] (4) -- node[midway, below] {1} (6);
\draw [->] (6) to [out=280,in=260] node[midway, below] {1} (4);
\draw [->] (5) -- node[midway, above right] {1} (6);
\draw [->] (6) to [out=40,in=25] node[midway, above right] {1} (5);
\draw [->] (4) -- node[midway, above left] {1} (5);
\draw [->] (5) to [out=140,in=140] node[midway, above left] {1} (4);

\draw [->] (7) -- node[midway, above right] {1} (8);
\draw [->] (8) to [out=30,in=45] node[midway, above right] {1} (7);
\draw [->] (8) -- node[midway, below right] {1} (9);
\draw [->] (9) to [out=320,in=320] node[midway, below right] {1} (8);
\draw [->] (9) -- node[midway, left] {1} (7);
\draw [->] (7) to [out=210,in=140] node[midway, left] {1} (9);

\draw [->] (3) to [out=10,in=170] node[midway, below] {5} (9);
\draw [->] (9) to [out=210,in=330] node[midway, below] {5} (3);

\draw [->] (2) to [out=50,in=180] node[midway, above] {5} (4);
\draw [->] (4) to [out=225,in=0] node[midway, below] {5} (2);

\draw [->] (6) to [out=290,in=190] node[midway, below] {5} (7);
\draw [->] (7) to [out=140,in=0] node[midway, above] {5} (6);

\node[main] (10) [above of=5, yshift=1.5cm, xshift=3.5cm]{$10$}; 
\node[main] (11) [above right of=10] {$11$};
\node[main] (12) [below right of=10] {$12$}; 
\node[main] (13) [above right of=11, xshift=1.5cm] {$13$};
\node[main] (14) [above right of=13] {$14$}; 
\node[main] (15) [below right of=14] {$15$};
\node[main] (16) [below right of=15, xshift=1.5cm] {$16$};
\node[main] (17) [below right of=16] {$17$}; 
\node[main] (18) [below left of=17] {$18$};

\draw [->] (12) -- node[midway, right] {1} (11);
\draw [->] (11) to [out=330,in=30] node[midway, right] {1} (12);
\draw [->] (10) -- node[midway, below left] {1} (12);
\draw [->] (12) to [out=230,in=255] node[midway, below left] {1} (10);
\draw [->] (10) -- node[midway, above left] {1} (11);
\draw [->] (11) to [out=100,in=170] node[midway, above left] {1} (10);

\draw [->] (13) -- node[midway, below] {1} (15);
\draw [->] (15) to [out=280,in=260] node[midway, below] {1} (13);
\draw [->] (14) -- node[midway, above right] {1} (15);
\draw [->] (15) to [out=40,in=50] node[midway, above right] {1} (14);
\draw [->] (13) -- node[midway, above left] {1} (14);
\draw [->] (14) to [out=140,in=140] node[midway, above left] {1} (13);

\draw [->] (16) -- node[midway, above right] {1} (17);
\draw [->] (17) to [out=30,in=45] node[midway, above right] {1} (16);
\draw [->] (17) -- node[midway, below right] {1} (18);
\draw [->] (18) to [out=320,in=320] node[midway, below right] {1} (17);
\draw [->] (18) -- node[midway, left] {1} (16);
\draw [->] (16) to [out=210,in=140] node[midway, left] {1} (18);

\draw [->] (12) to [out=10,in=170] node[midway, below] {5} (18);
\draw [->] (18) to [out=210,in=330] node[midway, below] {5} (12);

\draw [->] (11) to [out=50,in=180] node[midway, above] {5} (13);
\draw [->] (13) to [out=225,in=0] node[midway, below] {5} (11);

\draw [->] (15) to [out=290,in=190] node[midway, below] {5} (16);
\draw [->] (16) to [out=140,in=0] node[midway, above] {5} (15);

\draw [->] (5) to [out=90,in=190] node[midway, above] {10} (10);
\draw [->] (10) to [out=230,in=50] node[midway, above] {10} (5);

\end{tikzpicture}}
\caption{Example of nine node weighted directed graph used to build a more complex theoretical network.} \label{fig:ExampleMNnode}
\end{figure}

To simulate the theoretical networks used in the results we use the nine-node example in Figure \ref{fig:Example9node} as a building block. The single undirected edges of the simple graph are replaced with edges in both directions which are assigned equal distances. The process adds another of the nine-blocks to the network connecting a random node on the existing network to a random node on the new nine-node block. The random chosen nodes are limited to the nodes with order less than 6 (in and out combined). In the example of the process in Figure \ref{fig:ExampleMNnode} this restricts the connections to nodes 1, 5 and 8. A larger distance for the dual edges connecting the blocks than those within the nine-node unit is used.

We chose to create the networks with this approach as it represents a suitable approximation of how conurbations connect together and it contains a visible modular structure amenable to the methods applied.

A number of network blocks are connected to make the size of test network required (in multiples of nine). Once the network is specified, an O-D matrix is created for the network which randomly assigns a number between 0 and 10 to each O-D pair. The network is taken to be uncongested so the congestion function used is just the edge distance (independent of flow).

With the assigned O-D matrix the average flows on the network are created by using the Frank-Wolfe algorithm to solve for user-equilibrium (Eq. \ref{eq:UserObj}). This provides an average flow on each edge which can be used to generate a sample number of days of flows by using a random Poisson generator. The number of simulated days is set to be the number of edges in the network multiplied by 2.5. This flow sample is then used in the same processes described in the methodology to generate results.

For each size of network three iterations were trialled. The random aspect to the network creation did not have a considerable effect on the results.